\documentclass[twocolumn,resetfootnote]{aastex7}

\usepackage{lineno}
\linenumbers

\usepackage{graphicx}	% Including figure files
\usepackage{amsmath}	% Advanced maths commands

\usepackage[dvipsnames]{xcolor}
\usepackage{xspace}
\usepackage{anyfontsize}
\usepackage{natbib}
\usepackage{comment}
\usepackage{hyperref}

\newcommand{\decode}{\textsc{decode}\xspace}

\defcitealias{fu_2025a}{F25}
\defcitealias{fu_2026}{F26}

\definecolor{RezaRefereeGreen}{HTML}{006400}

\begin{document}

%\title{Galaxy merger rates at redshift $z\gtrsim5$: a comprehensive study from models and simulations}
\title{Mergers may not be the main channel for galaxy stellar mass growth at redshift $z\gtrsim5$:\\ Predictions from cosmological hierarchical models and simulations}

\author[orcid=0009-0002-8051-1056]{Hao Fu}
\affiliation{Center for Astronomy and Astrophysics, Department of Physics, Fudan University, Shanghai 200438, China}
%\affiliation{School of Physics and Astronomy, University of Southampton, Highfield, Southampton, SO17 1BJ, UK}
\email[show]{haofu@fudan.edu.cn}

\author[orcid=0000-0003-3783-2321]{Mohammadreza Ayromlou}
\affiliation{Argelander-Institut f\"ur Astronomie, Auf dem H\"ugel 71, D-53121 Bonn, Germany}
\email{}

\author[orcid=0000-0001-8973-5051]{Francesco Shankar}
\affiliation{School of Physics and Astronomy, University of Southampton, Highfield, Southampton, SO17 1BJ, UK}
\email{}

%\author[]{Lars Hernquist}
%\affiliation{Center for Astrophysics $\mid$ Harvard $\&$ Smithsonian, 60 Garden Street, Cambridge, MA 02138, USA}
%\email{}

\author[orcid=0000-0002-4882-1735]{Andrea Lapi}
\affiliation{SISSA, Via Bonomea 265, 34136 Trieste, Italy}
\affiliation{IFPU, Via Beirut 2, 34014 Trieste, Italy}
\affiliation{INFN, Sezione di Trieste, via Valerio 2, 34127 Trieste, Italy}
\affiliation{IRA-INAF, Via Gobetti 101, 40129 Bologna, Italy}
\email{}

%\author[orcid=0000-0003-0685-3525]{Rongrong Liu}
%\affiliation{Center for Astrophysics $\mid$ Harvard $\&$ Smithsonian, 60 Garden Street, Cambridge, MA 02138, USA}
%\email{}

\author[orcid=0009-0009-7662-0445]{Daniel Roberts}
\affiliation{School of Physics and Astronomy, University of Southampton, Highfield, Southampton, SO17 1BJ, UK}
\email{}

\author[]{Feng Yuan}
\affiliation{Center for Astronomy and Astrophysics, Department of Physics, Fudan University, Shanghai 200438, China}
\email{}

\author[orcid=0000-0002-0287-7485]{Maulik Bhatt}
\affiliation{Scuola Normale Superiore, Piazza dei Cavalieri 7, I-56126 Pisa, Italy}
\email{}

\author[orcid=0000-0003-2047-3684]{Evgenii Chaikin}
\affiliation{Institute for Computational Cosmology, Department of Physics, Durham University, South Road, Durham, DH1 3LE, UK}
\email{}

\author[orcid=0000-0002-7200-8293]{Simona Gallerani}
\affiliation{Scuola Normale Superiore, Piazza dei Cavalieri 7, I-56126 Pisa, Italy}
\email{}

\author[orcid=0000-0002-9231-1505]{Luca Graziani}
\affiliation{Dipartimento di Fisica, Sapienza, Universit\`a di Roma, Piazzale Aldo Moro 5, I-00185 Roma, Italy}
\affiliation{INFN, Sezione di Roma I, Piazzale Aldo Moro 2, I-00185 Roma, Italy}
\affiliation{Istituto Nazionale di Astrofisica (INAF), Osservatorio Astronomico di Roma, Via Frascati 33, 00078 Monte Porzio Catone, Italy}
\email{}

\author[]{Antonio La Marca}
\affiliation{European Space Agency/ESTEC, Keplerlaan 1, 2201 AZ Noordwijk, The Netherlands }
\affiliation{Leiden Observatory, Leiden University, Einsteinweg 55, 2333 CC Leiden, The Netherlands}
\email{}

\author[orcid=0000-0002-8660-6809]{George Mountrichas}
\affiliation{Instituto de Fisica de Cantabria (CSIC-Universidad de Cantabria), Avenida de los Castros, 39005 Santander, Spain}
\email{}

\author[orcid=0000-0003-0939-9671]{Yingjie Peng}
\affiliation{Department of Astronomy, School of Physics, Peking University, Beijing 100871, China}
\affiliation{Kavli Institute for Astronomy and Astrophysics, Peking University, Beijing 100871, China}
\email{}

\author[]{Haocheng Wang}
\affiliation{Center for Astronomy and Astrophysics, Department of Physics, Fudan University, Shanghai 200438, China}
\email{}

\author[orcid=0000-0002-3775-0484]{Kai Wang}
\affiliation{Institute for Computational Cosmology, Department of Physics, Durham University, South Road, Durham, DH1 3LE, UK}
\affiliation{Centre for Extragalactic Astronomy, Department of Physics, Durham University, South Road, Durham DH1 3LE, UK}
\email{}

%\author[]{}
%\affiliation{}
%\email{}

%\author{et al.}
%\affiliation{Center for Astronomy and Astrophysics, Department of Physics, Fudan University, Shanghai 200438, China}
%\email{}

%\author[orcid=0000-0000-0000-0002,gname=Bosque, sname='Sur America']{Forrest Sur Am\'{e}rica}
%\altaffiliation{Las Campanas Observatory}
%\affiliation{Universidad de Chile, Department of Astronomy}
%\email{fakeemail2@google.com}

%\collaboration{all}{The Terra Mater collaboration}

\begin{abstract}

Mergers are a crucial channel of galaxy growth, but their role relative to star formation remains uncertain at high redshift. Here we investigate merger rates and merger-driven stellar mass assembly for galaxies with stellar mass $M_\star \gtrsim 10^8 \, {\rm M_{\odot}}$ at $z\simeq 5-11$, by combining complementary theoretical approaches within $\Lambda$CDM. These include the DECODE semi-empirical model, the TNG100 and EAGLE hydrodynamical simulations, and the \textsc{L-Galaxies} semi-analytical model, together with the Millennium and TNG100-Dark simulations used to examine dark matter subhalo statistics and merger timescales. This compares models with different treatments of baryonic physics and galaxy-halo connections. We find that dark matter-only simulations predict merger timescales $\sim 30\%$ longer than those inferred from cosmological simulations including baryons. Galaxy major merger ($M_{\rm \star,sat} / M_{\rm \star,cen} > 0.25$) rates show a substantial model- and mass-dependent redshift evolution. Despite high rates, mergers contribute only $\lesssim 10\%$ to the total cosmic stellar mass density evolution. The average ex-situ stellar mass fraction ranges from a few percent to $\sim 25\%$ for all stellar mass bins at $z\gtrsim 5$, across different models and simulations. Their broad agreement implies that stellar mass assembly at cosmic dawn is dominated by in-situ star formation rather than mergers. These findings suggest that the high pair fractions identified by JWST may not directly trace efficient ex-situ stellar mass growth, highlighting the need for physically motivated merger timescales to interpret galaxy assembly in the early Universe.

\end{abstract}



\section{Introduction}\label{sec:intro}

Significant progress has been made in the modelling of galaxy formation and evolution within the Lambda Cold Dark Matter ($\Lambda$CDM) framework (e.g., \citealt{white_1991, kauffmann_1993, cole_2000}). However, some physical processes involved therein are still debated or not fully understood. For instance, it is still unclear to what extent galaxies acquire their stellar mass \textit{ex-situ} from mergers, against \textit{in-situ} processes like star formation (e.g., \citealt{guo_2008, cattaneo_2011, rodriguez_gomez_2016, pillepich_2018, davison_2020, fu_2022}). In a $\Lambda$CDM Universe galaxies are modelled within dark matter haloes with their evolution being closely connected. In particular, mergers with nearby smaller galaxies are thought to follow mergers between the host dark matter haloes, which are frequent in a hierarchical dark matter-dominated framework, after a merger timescale that depends on the dynamical friction (e.g., \citealt{lacey_cole_1993, fakhouri_2008, genel_2009, fakhouri_2010}). However, cosmological simulations showed that at low redshifts the merger timescales of galaxies are often longer than the Hubble time, leading to a satellite galaxy orbiting around a more massive central galaxy (e.g., \citealt{khochfar_2006, boylan_kolchin_2008, stewart_2009, hopkins_2010, mccavana_2012, zavala_2019, fu_2022}).

Towards high redshifts, the relative roles of mergers and star formation in contributing to the galaxy stellar mass growth become even less understood both from the observational and theoretical sides. However, the recent observations from the James Webb Space Telescope (JWST; \citealt{gardner_2006, gardner_2023}) have started to shed light on the merger rates of galaxies after the epoch of reionisation. In particular, studies of galaxy mergers have reported high major merger pair fractions at redshifts up to $z\sim 11$, but estimates of their contribution to the total stellar mass growth differ substantially, with \citet{duan_2025} suggesting that mergers account for up to half of this growth, whereas \citet{puskas_2025} and \citet{calabro_2026} inferring a contribution of $\lesssim10\%$. Although mergers were found to be frequent in traditional models, such large contributions from mergers to the stellar mass increase are still in contrast with the current findings in simulations, which suggest that the evolution at high redshift is mainly governed by star formation from cold gas with mergers becoming gradually important at lower redshifts in massive galaxies (e.g., \citealt{rodriguez_gomez_2016, pillepich_2018, davison_2020}), not only for the stellar mass growth but also for, e.g., the size evolution and morphology (e.g., \citealt{naab_2009, hopkins_2009, nipoti_2012, shankar_2013, shankar_2014, shankar_2015, rodriguez_gomez_2017}). These inconsistencies have also pushed the merger rates of current models to be revised, prompting speculation about shorter merger timescales or higher star formation efficiencies before infall.

From the observational side, the main methods available to study merger processes are the morphology-based selection and close-pair analysis. Morphology-based selection identifies galaxy mergers by analysing the shapes, structures, and visual disturbances in galaxy images, and is more powerful for detecting the post-merger phase (e.g., \citealt{conselice_2003, conselice_2008, lotz_2008, lotz_2010, whitney_2021, desmons_2023, dalmasso_2024}). The close pair-based selection identifies galaxy mergers by finding galaxies that are physically close to each other in projected distance, making it more efficient for detecting the pre-merger phase before galaxies show clear morphological disturbances (e.g., \citealt{le_fevre_2000, patton_2000, de_ravel_2009, lotz_2011, man_2016, mundy_2017, chaikin_2018, conselice_2022}). However, the above methods, as mentioned, allow us to observe galaxy pre- and post-mergers, whereas the predictions of the actual merger rates are typically based on dynamical friction timescales fitted from numerical simulations.

From the theoretical side, mergers have been widely studied in cosmological models and simulations (e.g., \citealt{hopkins_2010, hopkins2010b, rodriguez_gomez_2015, rodriguez_montero_2019, ferreira_2020, mcalpine_2020, oleary_2021, husko_2022, schechter_2025, chaikin_2026, lagos_2026}). Cosmological hydrodynamical simulations provide a self-consistent description of dark matter and baryonic evolution, and are therefore particularly powerful for measuring galaxy merger rates and timescales. However, their predictions may be affected by a number of numerical artifacts, such as the stellar mass cut-off, subhalo identification via the Friends-of-Friends algorithm (e.g., number of particles or radius definition), resolution effects, the implementation of the subgrid physics and the assumed parameters (e.g., \citealt{muldrew_2011, crain_2015, rodriguez_gomez_2015}). Semi-analytical models offer a complementary approach by following galaxy evolution along dark matter halo merger trees using physically motivated prescriptions (e.g., \citealt{kauffmann_1993, cole_2000, henriques_2015}), and semi-empirical models connect galaxies to their host haloes in a data-driven fashion through observationally constrained scaling relations with minimal assumptions about baryonic processes (e.g., \citealt{moster_2013, moster_2018, behroozi_2019}). Therefore, whilst hydrodynamical simulations explicitly model baryonic processes, semi-analytical and semi-empirical models offer greater flexibility for exploring parameter space and isolating the physical ingredients governing galaxy assembly. Together, these different frameworks allow us to assess the robustness of predicted merger histories against the specific treatment of galaxy formation physics and numerical implementation.

Here, we study galaxy merger histories using a variety of cosmological models and simulations. We will investigate at which rate galaxy mergers occur in the early Universe and what is the stellar mass that central galaxies acquire ex-situ. In particular, we use complementary theoretical frameworks, including the \decode (\citealt{fu_2025a, fu_2025b, fu_2026}) semi-empirical model, the TNG (\citealt{nelson_2018, pillepich_2018a}) and EAGLE (\citealt{schaye_2015}) cosmological hydrodynamical simulations, and the \textsc{L-Galaxies} (\citealt{henriques_2020, ayromlou_2021}) semi-analytical model. Together, these approaches provide independent yet complementary predictions for merger rates and ex-situ stellar mass growth within the $\Lambda$CDM paradigm. Our main goals are on one side to predict in a flexible and transparent way the contribution of mergers to galaxy stellar mass assembly at early epochs, and on the other side to understand the level of agreement or disagreement among different types of galaxy evolution models and approaches still embedded within the same cold dark matter cosmological framework.

\begin{table*}
    \caption{Observational data used in this work for comparison. Different columns show the work, survey name, redshift range, stellar mass coverage (with $m_\star \equiv \log_{10} \, (M_\star/M_\odot)$), survey area, and the quantity considered for comparison, where GMR represents the galaxy merger rate and SMD the stellar mass density.}
    \centering
    \begin{tabular}{cccccc}
        \hline
        Publication & Survey & Redshift range & Stellar mass & Survey area & Quantity \\
        \hline
        \citet{duncan_2019} & HST CANDELS & $0.5<z<6$ & $m_\star>9.7$ & $951\, {\rm arcmin}^2$ & GMR \\
        \citet{weaver_2023} & COSMOS2020 & $0.2<z<7.5$ & $m_\star\simeq8-12$ & $1.27\, {\rm deg}^2$ & SMD \\
        \citet{navarro_carrera_2024} & JWST HUDF \& UKIDSS & $3.5<z<8.5$ & $m_\star >8$ &
        $7-12 \, {\rm arcmin}^2$ & SMD \\
        \citet{weibel_2024} & JWST/NIRCam & $3.5<z<9.5$ & $m_\star\simeq 7.5-12$ & $500.8\, {\rm arcmin}^2$ & SMD \\
        \citet{duan_2025} & JWST Cycle-1 & $4.5<z<11.5$ & $8<m_\star<10$ & $189.36\, {\rm arcmin}^2$ & GMR \\
        \citet{puskas_2025} & JWST JADES & $3<z<9$ & $8<m_\star<10$ & $204\, {\rm arcmin}^2$ & GMR \\
        \citet{shuntov_2025} & JWST COSMOS-Web & $0.2<z<12$ & $m_\star\simeq 7.5-12$ & $0.43\, {\rm deg}^2$ & SMD \\
        \citet{calabro_2026} & JWST F444W & $5<z<14$ & $7<m_\star<11$ & $132\, {\rm arcmin}^2$ & GMR \\
        \hline
    \end{tabular}
    \label{tab:data}
\end{table*}

This paper is structured as follows. In Section \ref{sec:data}, we describe the observational datasets we compare our results to. In Section \ref{sec:models}, we describe the simulations and models that we use in this work. In Section \ref{sec:results}, we present our predictions for the dark matter halo merger rates and timescales, galaxy merger rates and ex-situ fractions. Finally, in Sections \ref{sec:discuss} and \ref{sec:conclusions}, we discuss our results and draw our conclusions. In what follows, we adopt the $\Lambda$CDM cosmology with slightly different cosmological parameters across models, although these do not alter the main results. All the input and reference data sets adopted here are based on a \citet{chabrier_2003} stellar initial mass function (IMF).

\section{Observational data}\label{sec:data}

In this paper we compare the predictions from the different models and simulations with observational constraints on both the galaxy merger rate and the cosmic stellar mass density (SMD) from the literature. The former provides a direct constraint on the frequency of merger events, whereas the latter traces the integrated build-up of the galaxy stellar mass allowing us to assess the relative contribution of mergers to the total stellar mass assembly. For the galaxy merger rates, we consider measurements based on both galaxy close pairs and merger-induced morphological disturbances, while for the cosmic SMD we use measurements derived from the galaxy stellar mass function (SMF). The datasets are summarised in Table \ref{tab:data} and described below.

For the galaxy merger rates, we adopt the measurements from \citet{duncan_2019}, \citet{duan_2025}, \citet{puskas_2025} and \citet{calabro_2026}. The former three apply a probabilistic close pair methodology based on the full photometric redshift probability distributions and select companions at different projected separation distances, redshifts and stellar masses. In particular, \citet{duncan_2019} identified galaxy pairs with stellar mass $M_\star \gtrsim 10^{9.7}\, M_\odot$ at projected separations of $5-30$ kpc out to $z\simeq6$, using the five fields from the Cosmic Assembly Near-infrared Deep Extragalactic Legacy Survey (CANDELS) of the Hubble Space Telescope. The fields cover approximately $951\,\mathrm{arcmin}^{2}$, with spatially varying $80\%$ detection-completeness limits of $H_{160}\simeq25.3–27.3$ mag. \citet{duan_2025} used eight JWST Cycle-1 fields to identify close pairs over $4.5<z<11.5$ for galaxies with $8<\log_{10}(M_\star/M_\odot)<10$, adopting a projected separation of $20-50$ kpc. The combined unmasked area is $189.36\,\mathrm{arcmin}^{2}$, with field-dependent $5\sigma$ F444W depths of $28.83–30.22$ mag. Similarly, \citet{puskas_2025} used deep JWST Advanced Deep Extragalactic Survey (JADES) imaging and available spectroscopy to measure close pair fractions at $z\simeq3-9$ for galaxies with $8<\log_{10}(M_\star/M_\odot)<10$, selecting companions within $5-30$ physical kpc. The medium and deep JADES cover approximately $204\,\mathrm{arcmin}^{2}$, with $5\sigma$ F444W point-source depths of $29.29–30.66$ mag. Finally, \citet{calabro_2026} provided an independent morphology-based estimate using a spectroscopic sample of galaxies at $5<z<14$ observed in six JWST/F444W fields. They identify mergers from rest-frame optical disturbances using non-parametric morphological indicators. The underlying F444W images reach $5\sigma$ depths of $\sim28–30$ AB mag, while the morphological analysis additionally requires an average signal-to-noise ratio per pixel $S/N_{\rm pixel}>3$. All these studies converted the observed pair fractions or morphological merger fractions into merger rates by adopting redshift-dependent merger observability timescales calibrated from cosmological simulations.

For the cosmic SMD, we compile the measurements from \cite{weaver_2023, navarro_carrera_2024, weibel_2024, shuntov_2025}, which are derived by integrating the observed galaxy SMFs. \citet{weaver_2023} measured the SMF and cosmic SMD at redshift $0.2<z\leq7.5$ using the COSMOS2020 catalogue to constrain the abundance of massive galaxies and reduce the impact of cosmic variance. The analysis covers an effective area of $1.27\,\mathrm{deg}^{2}$, reduced to $0.716\,\mathrm{deg}^{2}$ in the highest-redshift bin, $6.5<z\leq7.5$, and adopts a Spitzer/IRAC $3.6\,\mu\mathrm{m}$ magnitude limit of 26, corresponding approximately to $S/N\gtrsim5$. \citet{navarro_carrera_2024} used deep JWST observations of the Hubble Ultra Deep Field (HUDF) and UKIDSS Ultra Deep Survey field, supplemented with CANDELS data at the high-mass end, to constrain the SMF at $3.5\lesssim z \lesssim 8.5$ down to stellar masses of $\sim10^8 \, M_\odot$ at $z\simeq8$. The JWST observations cover approximately 12 and 7 $\mathrm{arcmin}^{2}$ in HUDF and PRIMER-UDS, respectively, reaching $5\sigma$ F444W depths of 28.10 and $27.80–27.95$ mag in $0.25''$-radius apertures. The complementary CANDELS catalogues cover approximately 170 and \(200\,\mathrm{arcmin}^{2}\) in GOODS-S and UDS, respectively. \citet{weibel_2024} combined JWST/Near Infrared Camera (NIRCam) observations from Cosmic Evolution Early Release Science Survey (CEERS), Public Release Imaging for Extragalactic Research (PRIMER) and JADES, covering an effective area of $500\; {\rm arcmin}^2$, with field-averaged $5\sigma$ F444W depths ranging from 28.21 to 29.20 mag, to construct rest-frame optically selected SMFs at $4\lesssim z \lesssim 9$ and infer the corresponding cosmic SMD. Finally, \citet{shuntov_2025} used COSMOS-Web, with an effective area of $\sim 0.43\,\mathrm{deg}^{2}$ after masking, to measure the SMF and trace the cosmic SMD over $0.2<z<12$. The NIRCam imaging reaches $5\sigma$ depths of $27.2–28.2$ mag in $0.15''$-radius apertures, while the analysis adopts $m_{\mathrm{F444W}}<27.5$, corresponding to approximately $80\%$ completeness in total magnitude.

\section{Numerical models and simulations}\label{sec:models}

In this work we make use of the \decode semi-empirical model, TNG100 and EAGLE hydrodynamical simulations, as well as the \textsc{L-Galaxies} semi-analytical model, to study galaxy merger histories at high redshifts. In addition we also make use of the Millennium, Millennium-II and TNG100-Dark dark matter-only (DMO) simulations to calibrate the merger timescales of infalling satellite subhaloes by comparing the predicted abundances of surviving subhaloes with those actually retrieved from the simulation, which will be used as input in the \decode modelling. The full list of the aforementioned models and simulations can be found in Table \ref{tab:models}. Below, we provide more details on each of these simulations.

    \subsection{The DECODE semi-empirical model}\label{sec:decode}

    We use the \decode (\citealt{fu_2026_DECODE}) semi-empirical model, as presented in \citet[][hereafter referred to as F26]{fu_2026}. \decode grows galaxies along their dark matter main progenitor branches by integrating in time the star formation rate (SFR) which is assigned at each time step via an input scaling relation between SFR and halo accretion rate (HAR), derived via abundance matching. Galaxies can then acquire further stellar mass via mergers with satellite galaxies, whose stellar mass is assigned via the stellar mass-halo mass (SMHM) relation produced for the central counterparts.

    \begin{table*}
        \caption{Numerical models and simulations considered in this paper.}
        \centering
        \begin{tabular}{cccc}
            \hline
            Model/simulation name & Type & Volume size per dimension & Aim \\
            \hline
            Millennium & Dark matter-only simulation & 714 Mpc & Calibration \\
            Millennium-II & Dark matter-only simulation & 143 Mpc & Calibration \\
            DECODE & Semi-empirical model & 350 Mpc & Prediction \\
            TNG100-Dark & Dark matter-only simulation & 110 Mpc & Calibration \\
            TNG100 & Hydrodynamical simulation & 110 Mpc & Prediction \\
            EAGLE (L100N1504) & Hydrodynamical simulation & 100 Mpc & Prediction \\
            \textsc{L-Galaxies} & Semi-analytical model & 700 Mpc & Prediction \\
            \hline
        \end{tabular}
        \label{tab:models}
    \end{table*}

    \subsubsection{Dark matter halo merger tree}\label{sec:DM}

    \decode is run on a catalogue of dark matter haloes with mass\footnote{Halo masses are defined as $M_{\rm 200c}$, i.e., the total mass enclosed within a spherical radius where the average density equals 200 times the critical density of the Universe, consistently with the definition adopted in the other simulations.} above $M_{\rm h} > 10^{10.5} \, {\rm M_{\odot}}$ randomly extracted from the \citet{tinker_2008} halo mass function, which is equivalent to different cosmological volume sizes at each redshift of interest. The mass accretion history of each dark matter halo and its merger tree are computed via the SatGen code by \citet{jiang_2021}, based on the \citet{parkinson_2008} algorithm. Each of these parent dark matter haloes is characterised by the main progenitor branch along with the full merger tree, which includes the infall redshift of the satellite subhaloes and their mass accretion history before infall. Central galaxies are evolved self-consistently within the main progenitors following the recipe described below.

    \subsubsection{Abundance matching}\label{sec:AM}

    The input SFR-HAR relation is computed through the abundance matching between the observed SFR function and the theoretical HAR function following Equation 37 in \citet{aversa_2015} (see, e.g., \citealt{fu_2022}, \citealt{boco_2023}, \citealt{fu_2024}, \citealt{fu_2025a}, \citealt{fu_2025b} and \citealt{fu_2026}). This formalism eliminates any assumption on the functional form and adopts just one input parameter, the assumed Gaussian scatter in SFR at fixed HAR, for which we assume a constant value with mass and redshift of $0.4$ dex. As discussed in \citetalias{fu_2026}, the mass growth histories of galaxies do not noticeably change when varying the scatter in SFR at fixed HAR within reasonable values.

    The HAR function is directly extracted from the derivatives of the catalogue's halo mass growths, while the SFR function is derived by combining the most recently observed ultraviolet (UV) galaxy luminosity function (LF) at high redshifts (see references in \citetalias{fu_2026}) converted into SFRs via the calibration factor from \citet{madau_dickinson_2014} at fixed IMF. SFRs are then integrated across cosmic time, assuming a stellar mass loss rate of $1-\mathcal R$, with $\mathcal R = 0.45$ (e.g., \citealt{reimers_1975, renzini_1988, girardi_2000, pietrinferni_2004}). As detailed in \citetalias{fu_2026}, the simultaneous inclusion of dust and a variable IMF scenario can accordingly increase or decrease the implied star formation efficiencies while still generating similar star formation histories. Here we only consider the reference model with a fixed \citet{chabrier_2003} IMF, which can reproduce the measured SMFs and yield physically plausible star formation efficiencies\footnote{The star formation efficiency is defined as the efficiency in converting baryons into stars, i.e., the ratio between stellar mass and halo mass, as done in \citetalias{fu_2026}.}, to probe the galaxy merger rates.

    \subsubsection{Merging galaxies}\label{sec:method_merge_Gals}

    Satellite galaxies are initialised at infall on the mean SMHM relation characterising central galaxies in our mock catalogue at that epoch. As shown in \citet{fu_2022}, the shape of the SMHM relation directly impacts the galaxy merger rates, with flatter slopes producing more major mergers and vice versa. In \decode, the SMHM relation is not a free input, but it is directly implied by our input observed SFR functions via the time-integration of the star formation histories along the main progenitors.

    We assume satellites do not further grow or lose their stellar masses after infall via star formation or stripping, which is a reasonable approximation given the relatively short merger timescales at these high redshifts. Indeed, we showed in \citet{fu_2024} that the relative roles of star formation and stripping compensate each other, at least at lower redshifts, supporting the \textit{frozen} model as an initial approximation for the mean evolution of satellites after infall (see also \citealt{davies_2015}). We checked that even in the extreme scenario where all satellites continue forming stars at a rate given by the main sequence, the merger rates and ex-situ fractions would increase by a factor of up to $2$ in the less massive bins ($M_\star \simeq 10^9 \, M_\odot$) and do not alter appreciably in the most massive bins ($M_\star \gtrsim 10^{10} \, M_\odot$).

    We constrain the reliability of the frequency and duration of our dark-matter-only merger rates by comparing our predicted mass function of surviving unevolved subhaloes with those extracted from N-body simulations. A robust merger timescale would imply an accurate number of unmerged subhaloes. We compute the satellite subhalo merging timescale using Equation (5) of \citet{boylan_kolchin_2008},
    \begin{equation}\label{eq:tau_merge_boylan_kolchin}
    \begin{split}
        \tau_{\rm merge} = \tau_{\rm dyn} A \frac{ (M_{\rm h,cen} / M_{\rm h,sat})^b }{ \ln (1 + M_{\rm h,cen} / M_{\rm h,sat}) } & \\
        \exp \bigg[ c \frac{J}{J_{\rm c} (E)} \bigg] \bigg[ & \frac{r_{\rm c}(E)}{ R_{\rm vir} } \bigg]^d \, ,
    \end{split}
    \end{equation}
    which depends on the central-to-satellite halo mass ratio, the orbital circularity $J/J_{\rm c} (E)$, orbit radius $r_{\rm c}$, host halo virial radius $R_{\rm vir}$ and the free parameters $(A,b,c,d)$. The halo dynamical friction timescale $\tau_{\rm dyn}$ is computed following Equation (2) of \citet{jiang_2016}. The fitting parameters are reported in Section \ref{sec:res_halo_mergers}.

    %\rezac{Just randomly noticed this: Equation (\ref{eq:tau_merge_boylan_kolchin}) has $J/J_{\rm d}(E)$ while the text has $J/J_{\rm c}(E)$. Could we check which one is intended? Also, $J/J_{\rm c}(E)$ is the orbital circularity, i.e. angular momentum relative to a circular orbit of the same energy, not the orbital energy itself.}

    \subsection{The Millennium simulations}\label{sec:millennium}

    We make use of the Millennium simulations to constrain merger timescales and test the performance of \decode for predicting the abundance of dark matter subhaloes across cosmic time. Both the Millennium \citep{springel_2005} and Millennium-II \citep{boylan_kolchin_2009} simulations are rescaled to the Planck cosmology (\citealt{planck2018_cosmo_params}) following the method of \citet{angulo_white_2010}. The Millennium simulation covers a large cosmological volume ($714 \, {\rm Mpc}$ on a side) with a mass resolution of $1.43 \times 10^9 \, {\rm M_{\odot}}$, whereas Millennium-II covers a smaller volume ($143 \, {\rm Mpc}$ on a side) at a hundred times better mass resolution of $1.14 \times 10^7 \, {\rm M_{\odot}}$.

    Haloes are identified with the Friends-of-Friends (FOF) algorithm \citep{davis_1985}, and subhaloes with the SUBFIND algorithm \citep{springel_2001}. A subhalo must contain at least 20 particles to be resolved. This corresponds to a minimum resolved halo mass of $2.86 \times 10^{10} \, {\rm M_{\odot}}$ for the Millennium simulation, and $2.28 \times 10^8 \, {\rm M_{\odot}}$ for the Millennium-II simulation. Each halo contains one central subhalo, typically its most massive bound structure, while the remaining bound structures are classified as satellite subhaloes.

    In this work, we classify satellite subhaloes into two categories: surviving and disrupted. Surviving satellite subhaloes remain resolved by SUBFIND at the redshift of interest, whereas disrupted satellite subhaloes no longer do. In both cases, we track the subhalo mass until disruption and record its infall mass and infall redshift. Throughout this paper, the mass at infall refers to the $M_{\rm 200c}$ (hereafter $M_{\rm halo}$) mass of the satellite subhalo at the infall redshift, where $M_{\rm 200c}$ is the mass enclosed within a radius whose density is 200 times the critical density of the Universe at that redshift. This is the epoch when the satellite was last a central subhalo embedded in its own dark matter halo. For each host halo, we identify the surviving subhalo population and the full satellite population, including disrupted systems, and construct their mass functions at the redshift of interest.

    \subsection{The IllustrisTNG simulation}\label{sec:TNG}

    We also use data from the TNG100\footnote{\href{https://www.tng-project.org/}{https://www.tng-project.org/}} simulation, one component of the IllustrisTNG hydrodynamical simulation suite (hereafter TNG; \citealt{marinacci_2018, naiman_2018, nelson_2018, nelson_2019, pillepich_2018, springel_2018}). TNG100 has a box size of $110 \, {\rm Mpc}$ on a side, a dark matter mass resolution of $7.5 \times 10^6 \, {\rm M_{\odot}}$, and a baryonic mass resolution of $1.4 \times 10^6 \, {\rm M_{\odot}}$. The simulation is run with the AREPO moving-mesh code \citep{springel_2010} and includes subgrid prescriptions for the main processes that regulate galaxy formation, including gas cooling, star formation, stellar evolution, stellar feedback, and supermassive black hole seeding, growth, merging, and feedback \citep{weinberger_2017, pillepich_2018a}. Like the Millennium simulations, haloes and subhaloes in TNG are identified with the FOF and SUBFIND algorithms \citep{davis_1985, springel_2001}, respectively.

    TNG is calibrated against a broad set of observations of the galaxy population, including the cosmic SFR density, the SMF, and the stellar-to-halo mass ratio at $z=0$, while considering the black hole-stellar mass relation, halo gas fractions, and galaxy sizes. The calibration is performed at the fiducial TNG100 resolution and then the same model parameters were used in simulations with other resolutions, which can introduce non-negligible resolution-dependent effects \citep{pillepich_2018a, ayromlou_2021a}. We therefore focus on the calibrated version, TNG100, which provides a good compromise between mass resolution and statistical power. We use the TNG100 full run to study galaxy merger contributions at high redshift and the dark matter-only counterpart, TNG100-Dark, to study the statistics of dark matter subhaloes.

    For merger measurements, we use the publicly available subhalo merger trees constructed with the \textsc{sublink} algorithm \citep{rodriguez_gomez_2015, rodriguez_gomez_2016}. For each galaxy or subhalo selected at the redshift of interest, for example $z=5$ (${\rm snapshot}=17$), we trace the main progenitor branch backward through the simulation up to ${\rm snapshot}=2$ using the \texttt{FirstProgenitor} link. At each snapshot, additional progenitors are identified through the \texttt{NextProgenitor} links and are associated with the merging satellite population. This allows us to decompose the stellar mass growth of each galaxy into an in-situ component, formed through star formation along the main branch, and an ex-situ component, gained through mergers with smaller systems \citep{rodriguez_gomez_2016}.

    \subsection{The EAGLE simulation}\label{sec:EAGLE}

    In addition to TNG, we also use the EAGLE\footnote{\href{http://www.eaglesim.org/}{http://www.eaglesim.org/}} (Evolution and Assembly of GaLaxies and their Environments) simulation suite \citep{crain_2015, schaye_2015}. EAGLE is run with a modified version of the \textsc{gadget} code \citep{springel_2005_gadget}, using the conservative pressure-entropy formulation of smoothed particle hydrodynamics (SPH) derived by \citet{hopkins_2013_SPH}, the artificial viscosity switch from \citet{cullen_2010}, an artificial conduction switch similar to that of \citet{price_2008}, the \citet{wendland_1995} C2 kernel and the time-step limiters of \citet{durier_dallavecchia_2012}.

    As a hydrodynamical simulation, EAGLE models galaxy formation with subgrid prescriptions for the main baryonic processes, including radiative cooling, star formation, chemical enrichment, black hole growth, stellar feedback, and AGN feedback \citep{springel_2005b, springel_2003, booth_2009, wiersma_2009a, wiersma_2009b, dalla_vecchia_2012, rosas_guevara_2015, schaye_2015}. Haloes are identified with the FOF algorithm \citep{davis_1985}, and subhaloes with the SUBFIND algorithm \citep{springel_2001}.

    The reference EAGLE model is calibrated primarily to reproduce several key low-redshift observables, including the galaxy SMF, galaxy sizes, and the black hole mass-stellar mass relation (\citealt{crain_2015}). Here we use the publicly available merger trees from the largest reference run, Ref-L100N1504, which evolves a box size of $100\,{\rm Mpc}$ on a side. For each galaxy selected at the redshift of interest, for example $z\sim5$ (${\rm snapshot}=8$), we identify its descendant branch through the database links and reconstruct the progenitor population at earlier times up to ${\rm snapshot}=1$ using the \texttt{GalaxyID} and \texttt{DescendantID} fields. At a given snapshot, the progenitor with the largest stellar mass is taken to be the main progenitor, while the remaining progenitors are associated with merging satellites. This procedure allows us to separate the stellar mass assembled through in-situ star formation from that accreted ex-situ through mergers (\citealt{mcalpine_2016, qu_2017}).

    \subsection{The L-Galaxies semi-analytical model}\label{sec:Lgalaxies}

    In this work, we use the version of the \textsc{L-Galaxies}\footnote{\href{https://lgalaxiespublicrelease.github.io/}{https://lgalaxiespublicrelease.github.io/}} semi-analytical model for galaxy formation from \citet{ayromlou_2021}, which is built on top of the model of \citet{henriques_2020} and incorporates an improved recipe for gas stripping. \textsc{L-Galaxies} is a computationally efficient yet physically motivated model for simulating galaxy formation within the hierarchical growth of structures. It is built on dark matter halo merger trees extracted from the Millennium simulations and includes the main physical processes that regulate galaxy evolution, including gas accretion onto haloes, radiative cooling, star formation, stellar feedback, environmental effects, supermassive black hole growth, active galactic nucleus (AGN) feedback, and galaxy mergers.

    \begin{figure*}
        \centering
        \includegraphics[width=0.325\textwidth]{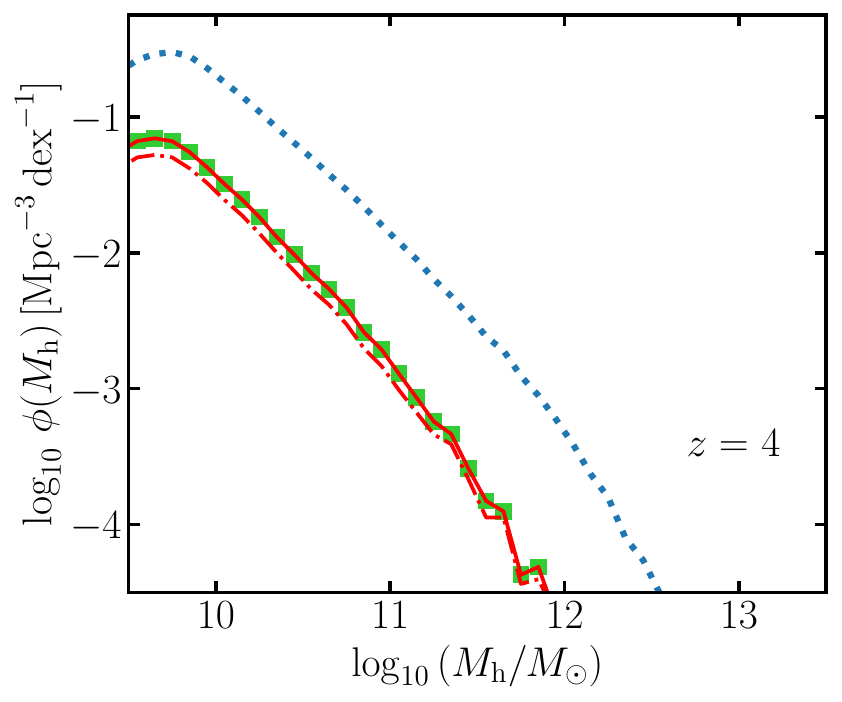}
        \includegraphics[width=0.325\textwidth]{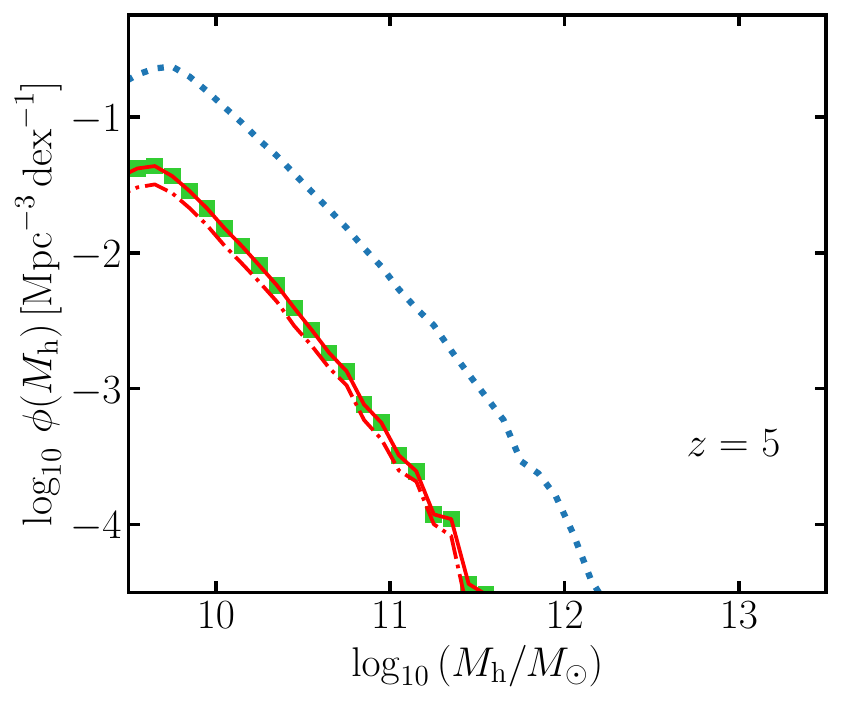}
        \includegraphics[width=0.325\textwidth]{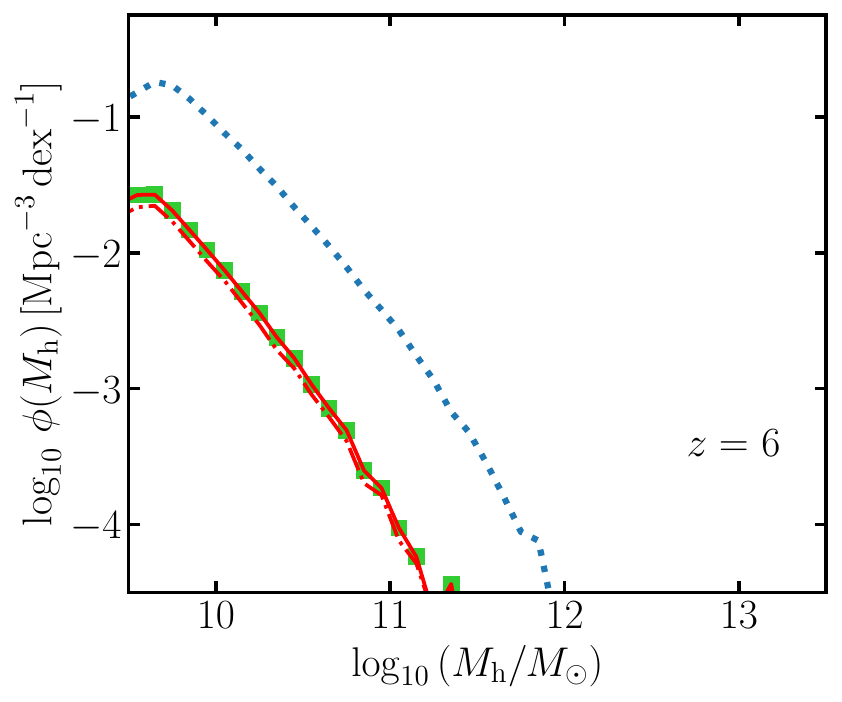}
        \includegraphics[width=0.325\textwidth]{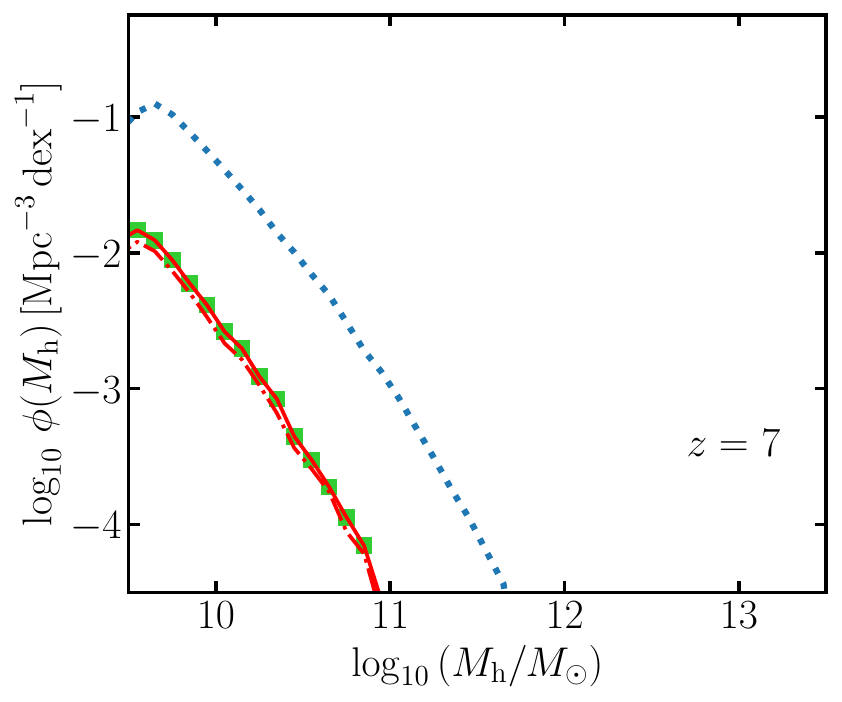}
        \includegraphics[width=0.325\textwidth]{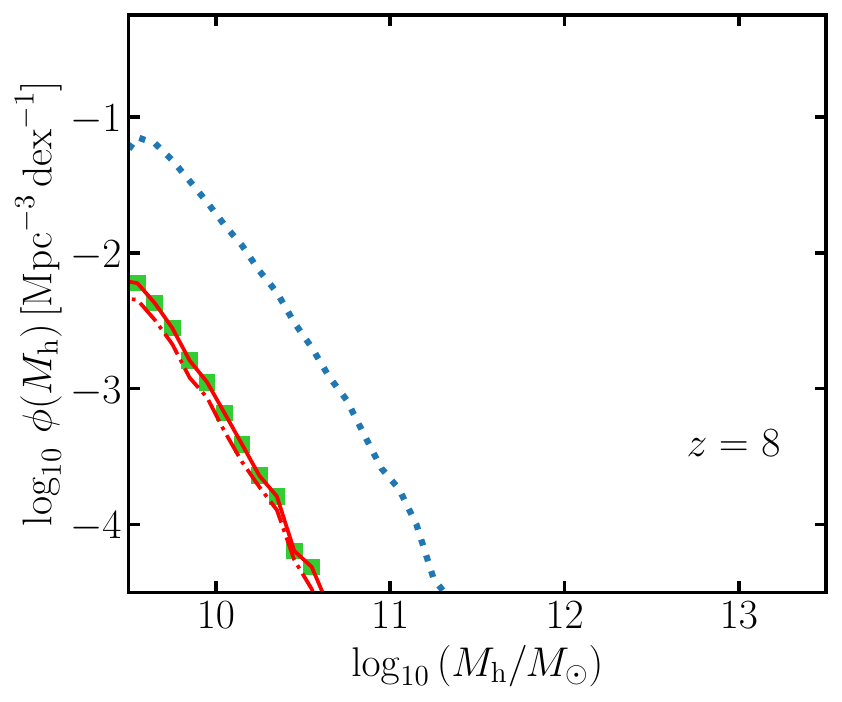}
        \includegraphics[width=0.325\textwidth]{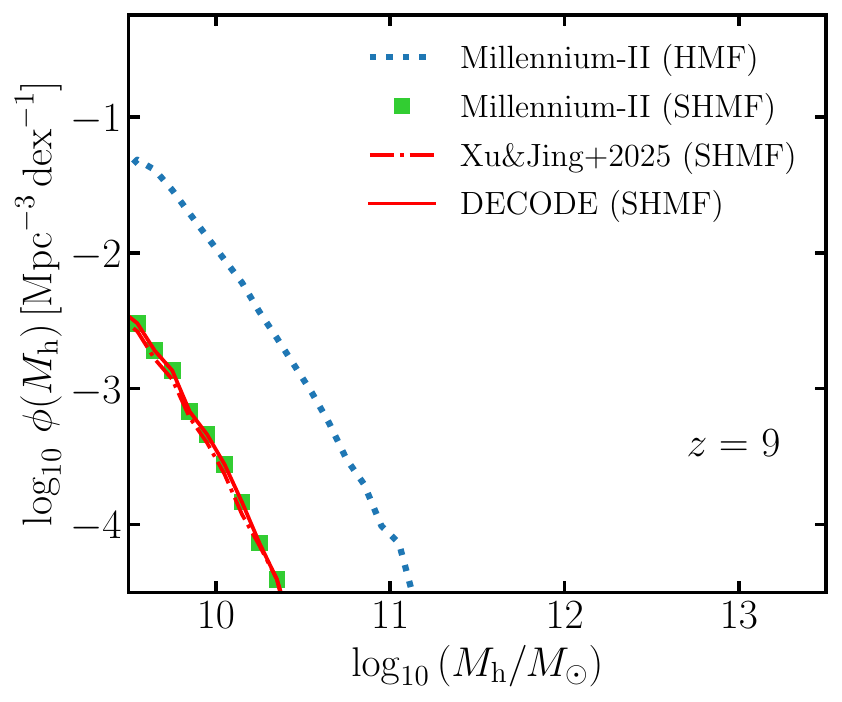}
        \caption{Halo mass function at different redshifts. The blue dotted lines and green squares show the mass function from the Millennium-II simulation for central subhaloes and surviving unevolved subhaloes, respectively. The red dash-dotted and solid lines show the surviving subhalo mass function computed using the merger timescales from \citet[][dark matter-only runs]{xu_jing_2025} and this work, respectively.}
        \label{fg:surv_shmf}
    \end{figure*}

    By following these processes within evolving dark matter haloes, \textsc{L-Galaxies} predicts a broad range of galaxy properties, including stellar masses in different components, SFRs, hot and cold gas content, morphology, black hole masses and accretion rates, and merger histories. The model is calibrated against several observational constraints, including the galaxy SMF and the quenched fraction over $z=0-2$, to reproduce the evolution of galaxy populations across cosmic time.

    Here we use the \textsc{L-Galaxies} realizations built on the Millennium \citep{springel_2005} and Millennium-II \citep{boylan_kolchin_2009} simulations to study galaxy merger histories at high redshift. In \textsc{L-Galaxies}, galaxies initially form within subhaloes and follow their positions and velocities. When a galaxy's host subhalo is no longer resolved, the galaxy is retained in the model rather than being removed. Its subsequent orbit is followed analytically by tracking the most bound particle of the disrupted subhalo, and the galaxy is allowed to merge after an assigned merger timescale, computed using the \citet{binney_1987} formula. This treatment provides a more realistic description of galaxy mergers after the subhalo disruption.

    The subhalo merger trees are constructed by linking subhaloes across simulation snapshots, which allows us to trace the formation and evolution of galaxies over cosmic time. For each galaxy, we identify its full progenitor set, including the main progenitor and secondary progenitors associated with merging satellites. When two galaxies merge, the main branch is defined as the one with the most massive history (e.g., \citealt{de_lucia_2007}). The merger trees therefore allow us to separate stellar mass growth from in-situ star formation and ex-situ accretion through mergers.

\section{Results}\label{sec:results}

In this section we present our results on the halo and galaxy mergers. We will start by presenting the level of performance of the \decode model in matching, with adequate choices of the dynamical friction timescales, the unevolved (unstripped) subhalo mass function (SHMF) as extracted from dedicated N-body simulations. Based on these timescales, we will then show the implied halo and thus galaxy merger rates, and the relative contribution of mergers in building up the stellar mass of high-redshift galaxies, along with the results from independent cosmological simulations and semi-analytical models.

    \subsection{Halo merger rates and timescales}\label{sec:res_halo_mergers}

    We here focus our attention on the subhalo population and their dynamical friction timescales. In Appendix \ref{app:shmf} we show that the unevolved total SHMF as computed from \decode's catalogue (described in Section \ref{sec:DM}) is consistent with the Millennium-II and TNG100-Dark simulations. We also note that, due to the limits in mass resolution, the statistics of the Millennium and TNG100-Dark rapidly become incomplete below halo mass ratios of $M_{\rm h,sat} / M_{\rm h,cen} \lesssim 0.3$ and $\lesssim0.25$, respectively. This would unavoidably limit their predictive power, especially for minor mergers, highlighting the need for analytic merger trees as a complementary approach. Indeed, we stress that an analytic approach to generate merger trees, validated against the outputs of N-body simulations within the resolution region of the latter, provides a powerful tool to scale up the generation of halo catalogues and assembly histories in number and mass resolution (see, e.g., Figure \ref{fg:total_unevolved_shmf}) with ease and extreme flexibility, without the need to rerun expensive simulations.

    We also note that the halo merger timescales in simulations can change depending on the presence or absence of baryons. For example, \citet{xu_jing_2025} showed that in the TNG100 and TNG300 simulations, the runs with baryons produce slightly shorter timescales with respect to the dark matter-only runs for low-mass haloes ($M_{\rm h} \lesssim 10^{12.5}\, M_\odot$), and vice versa in high-mass haloes. The presence of baryons shortens the merger timescales by $\sim 30\%$ (as discussed below). However, such a moderate shift in the merger timescales has a relatively moderate impact on our results on galaxy mergers, as further detailed in the next sections.

    Figure \ref{fg:surv_shmf} shows the unevolved surviving SHMF between redshift $z=4-9$. The unevolved surviving SHMF is defined as the number density of the satellite subhaloes, with their mass taken at infall, that have not yet merged at a given redshift $z$ (see also \citealt{jiang_2014, jiang_2016, van_den_bosch2016}). The surviving subhaloes represent an accurate test, not only for the overall statistics, but also for the timescale at which subhaloes merge. The green squares and red lines compare the SHMF of surviving subhaloes computed from the Millennium-II simulation to that obtained applying merger timescales from our fit and \citet{xu_jing_2025} to the Millennium-II subhaloes. The blue dotted lines show the mass function of central subhaloes from the Millennium-II.

    Our timescales, when applied to the total subhalo population of the Millennium-II, can well match the population of surviving subhaloes of the simulation itself at all redshifts above $z\geq 4$. We fit the parameters in Equation (\ref{eq:tau_merge_boylan_kolchin}) to reproduce the outputs of the N-body simulation, finding the following best-fitting parameters:
    \begin{equation}
        A=0.885\,, \;\; b=0.995\,, \;\; c=0.602\,, \;\; d=0.101 \,.
    \end{equation}

    \begin{figure}
        \centering
        \includegraphics[width=\columnwidth]{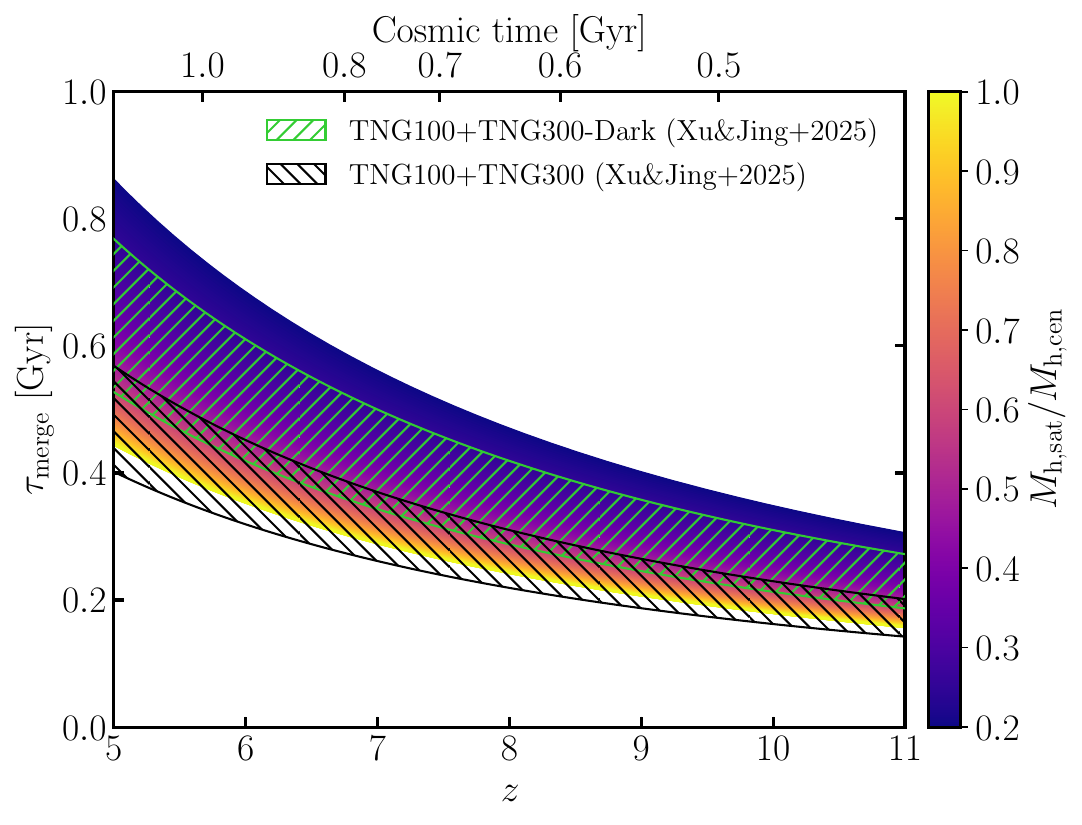}
        \caption{%Upper panel:
        Redshift evolution of merger timescale as a function of mass ratio. Timescales of \decode as a function of the mass ratio between the subhalo and the central halo are represented by the colour-coded area. The green slashed and black back-slashed areas show the timescales fit from \citet{xu_jing_2025} for $M_{\rm h} \sim 10^{12}\, M_\odot$, computed combining the TNG100 and TNG300 data from the dark matter-only and baryonic runs, respectively.}
        \label{fg:tau_merge}
    \end{figure}

    Figure \ref{fg:tau_merge} shows the merger timescales, for which we assume an average eccentricity of $0.5$, as a function of redshift used in this work. For completeness, in Figure \ref{fg:tau_merge} we also include the timescales from \citet{xu_jing_2025} extracted from the TNG100 and TNG300 dark matter-only runs. Reassuringly, we find that the timescales from the TNG dark matter-only simulations are in good agreement with our estimates from the Millennium simulations (green slashed and colour-coded regions, respectively). The TNG with baryons instead predicts $\sim 30\%$, on average, lower merger timescales (black back-slashed area) for the halo mass range of interest in this work (i.e., $M_{\rm h} \lesssim 10^{12.5}\, M_\odot$). Such a systematic decrease in dynamical friction timescales when including baryons has a relatively mild effect on the number of surviving subhaloes, as reported in Figure \ref{fg:surv_shmf}, and implied merger rates, as further discussed in the next sections.

\begin{figure}
        \centering
        \includegraphics[width=\columnwidth]{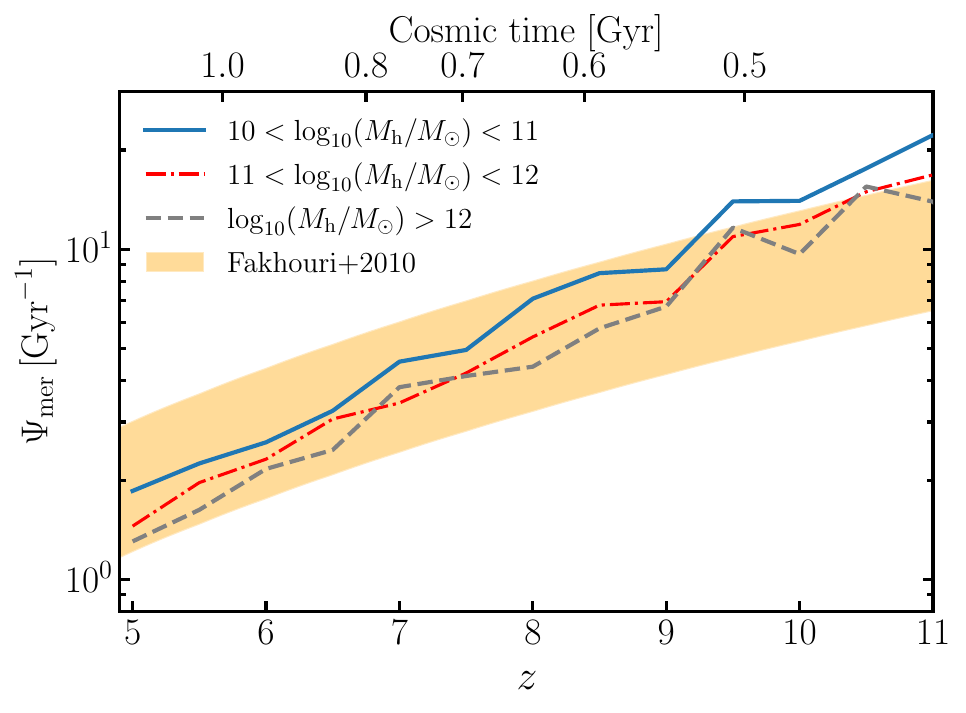}
        \caption{Halo major merger rate (number of mergers per unit time per halo with $M_{\rm h,sat}/M_{\rm h,cen} > 0.25$) as a function of redshift, for different halo mass bins. The blue solid, red dash-dotted and grey dashed lines show our predicted mean merger rate for central haloes in different halo mass bins, as labelled. The orange shaded area shows the merger rate from the Millennium simulation, as presented in \citet{fakhouri_2010}, in the halo mass range of $10^{10} < M_{\rm h} /M_\odot <10^{13}$.}
        \label{fg:DMhalo_merger_rate}
    \end{figure}

    Lastly, Figure \ref{fg:DMhalo_merger_rate} shows the halo-halo major merger ($M_{\rm h,sat}/M_{\rm h,cen} > 0.25$) rate. We find that the merger rate of dark matter haloes is relatively independent of the central halo mass from redshift $z\sim 5$ to $z\sim 11$, similar to the findings from \citet{fu_2022} at lower redshifts, and steadily increasing with redshift. Moreover, our merger rates are consistent with those from the Millennium simulations, as shown in \citet{fakhouri_2008} and \citet{fakhouri_2010}.

    \subsection{Galaxy merger rates}

    We now present the galaxy merger rates predicted by the complementary theoretical tools considered in this work. For \decode, the galaxy merger rates follow from the halo merger histories and the SMHM relation that is gradually built in time from the integration of the star formation histories. These predictions depend primarily on two input assumptions, namely the merger timescales fitted to dark matter-only simulations and the growth of the satellite galaxies after infall. The corresponding merger rates from TNG100, EAGLE and \textsc{L-Galaxies} are instead extracted directly from their merger trees, allowing us to assess the agreement among approaches with different treatments of baryonic physics and galaxy evolution.

    %\rezar{Regarding the correction described in your reply: could we state here exactly how the mean merger rate is now calculated, including galaxies with zero mergers (do they even exist? or are they very rare?)?}

    \begin{figure}
        \centering
        \includegraphics[width=\columnwidth]{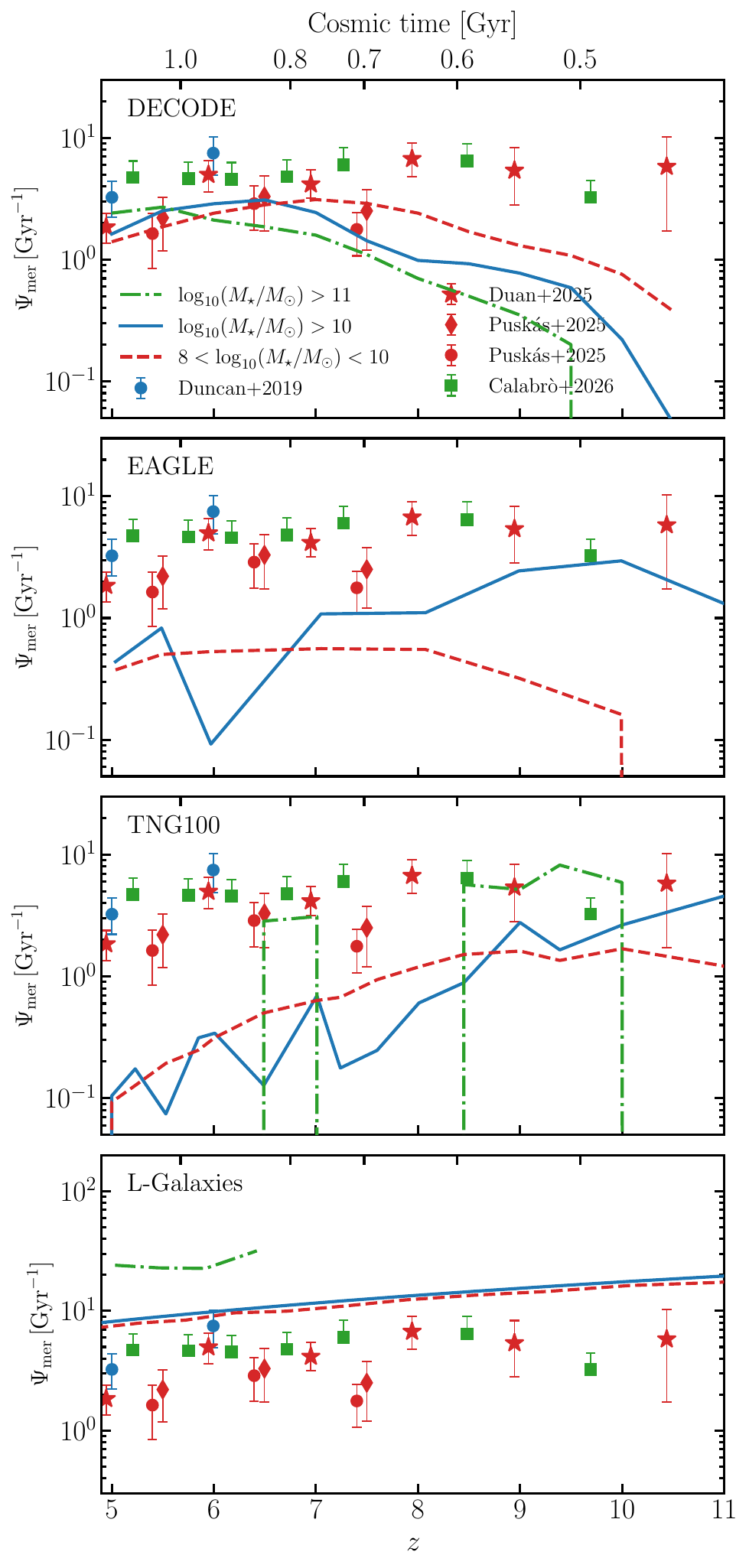}
        \caption{Galaxy major merger rate (number of mergers per unit time per galaxy with $M_{\rm \star,sat}/M_{\rm \star,cen} > 0.25$) as a function of redshift for different central galaxy stellar mass bins, as predicted by \decode (upper panel), EAGLE (second panel), TNG100 (third panel) and \textsc{L-Galaxies} (lower panel). The green dash-dotted, blue solid and red dashed lines show our predicted mean merger rate for central galaxies of stellar mass $\log_{10}(M_\star/{\rm M_{\odot}})>11$, $\log_{10}(M_\star/{\rm M_{\odot}})>10$ and $8<\log_{10}(M_\star/{\rm M_{\odot}})<10$, respectively. The data points with error bars show the merger rates from different observations of galaxy pairs in the same stellar mass bins as coloured (\citealt{duncan_2019}, \citealt{duan_2025}, \citealt{puskas_2025} and \citealt{calabro_2026}).}
        \label{fg:majormerger_rates_perGal}
    \end{figure}

    Figure \ref{fg:majormerger_rates_perGal} shows the galaxy major merger ($M_{\rm \star,sat} / M_{\rm \star,cen} > 0.25$) rate, expressed as the number of mergers per unit time, as a function of redshift for different stellar mass bins. We compare predictions from \decode, TNG, EAGLE and \textsc{L-Galaxies} to the merger rates from different data sets, estimated combining observed pair fractions or morphology-based merger fractions with timescales from cosmological simulations (\citealt{duncan_2019, duan_2025, puskas_2025, calabro_2026}), as described in Section \ref{sec:data}. We note that the timescales adopted therein are calculated as the merger observability timescales, i.e., since the time when a galaxy pair has a separation distance of $<30 \, {\rm kpc}$, rather than the infall time of the host subhalo (e.g., \citealt{snyder_2017, conselice_2022}). This difference in definition naturally produces different timescales but does not hinder the comparison, since the observability timescales are applied to observed close pairs, whereas our merger timescales are applied to infalling subhaloes. Therefore, both methods ultimately infer the corresponding galaxy merger rates.

    Merger rates in \decode are relatively consistent in normalisation and redshift dependence with the results from pair fraction estimates at $z\lesssim7.5$ and show a mild dependence on the central galaxy stellar mass bin. We stress that \decode's merger rates are a direct by-product of the underlying mapping between stellar mass and halo mass, which in turn is derived from the time integration of the input empirical SFRs, and of the merger timescales fine-tuned to reproduce the unevolved surviving SHMF. The more massive central galaxy stellar mass bin is, the stronger is the dependence on redshift, reporting a lower number of major mergers at earlier epochs, due to the steepening and evolution in slope of the SMHM relation at $z \gtrsim 7-8$. This dependence is also in agreement with the semi-analytic findings from \citet{husko_2022} and determinations from \citet{puskas_2025}. On the other hand, merger rates from the TNG and EAGLE simulations are slightly lower in normalisation and increase gradually towards higher redshifts. Lastly, \textsc{L-Galaxies} predicts merger rates approximately one order of magnitude higher than those from other models and observed pair fractions, which translates into a higher ex-situ stellar mass fraction, as we will show in Section \ref{sec:res_exsitu}. The reason behind the different major merger rates may reside in the slightly different mapping between stellar mass and halo mass, as tiny alterations in the latter can impact noticeably the galaxy major merger rates at fixed halo merger rate (see, e.g., \citealt{grylls_2019, fu_2022}). Additionally, the fact that in \textsc{L-Galaxies} orphan satellite galaxies can continue to merge, instead of being disrupted with their host haloes, can further contribute to enhancing the merger rates.

    \begin{figure}
        \centering
        \includegraphics[width=\columnwidth]{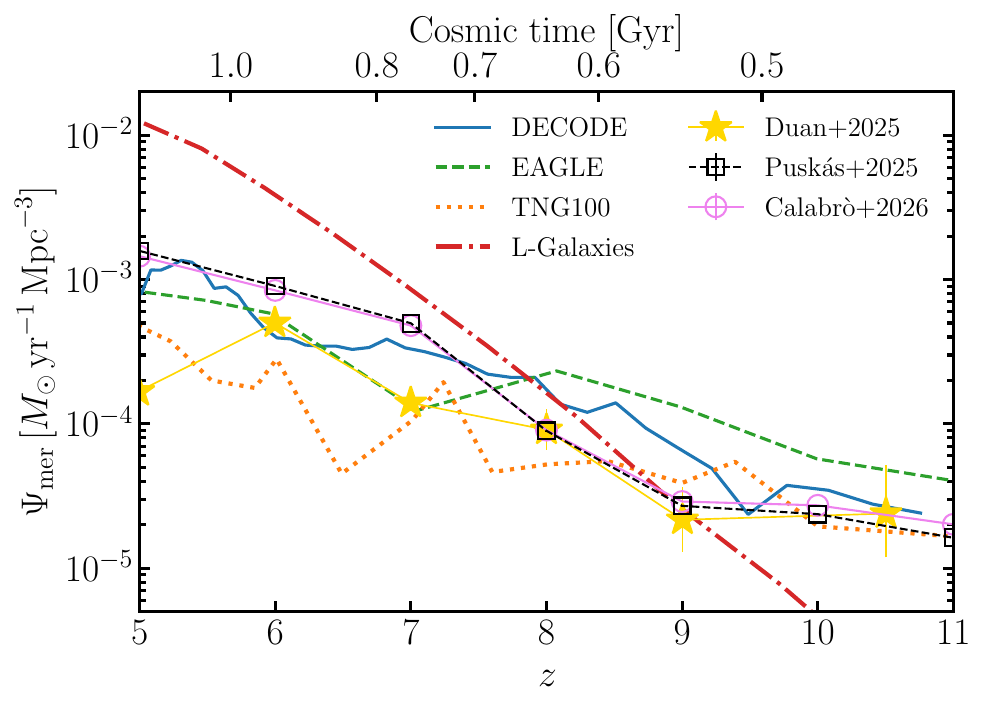}
        \caption{Total stellar mass accretion rate density from major mergers ($M_{\rm \star,sat} / M_{\rm \star,cen} > 0.25$) as a function of redshift, for central galaxies with stellar mass $10^8 \, {\rm M_{\odot}} < M_\star < 10^{10} \, {\rm M_{\odot}}$. The blue solid, green dashed, orange dotted and red dash-dotted lines show the prediction from \decode, EAGLE, TNG100 and \textsc{L-Galaxies}, respectively. The data points show the results from \citet[][yellow stars]{duan_2025}, and reconstructed estimates from \citet[][black squares]{puskas_2025} and \citet[][pink circles]{calabro_2026} based on observed pair fractions and morphology.}
        \label{fg:merger_rates}
    \end{figure}

    \begin{figure}
        \centering
        \includegraphics[width=\columnwidth]{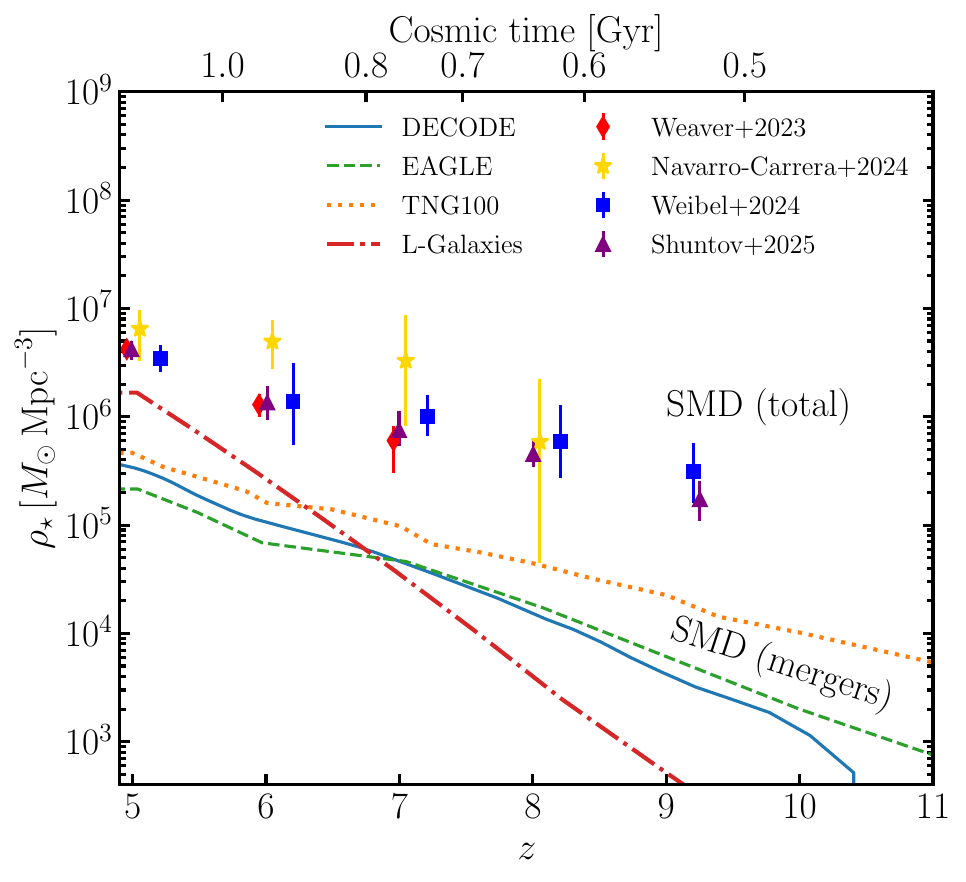}
        \caption{Integrated stellar mass density from mergers (ex-situ-only contribution from both minor and major mergers) for galaxies with stellar mass $M_\star > 10^8 \, M_\odot$, as predicted by \decode (blue solid line), EAGLE (green dashed line), TNG100 (orange dotted line) and \textsc{L-Galaxies} (red dash-dotted line). Data points show the cosmic total stellar mass density evolution from COSMOS2020 (red rhombuses with error bars; \citealt{weaver_2023}), JWST HUDF and UDS fields (yellow stars with error bars; \citealt{navarro_carrera_2024}), JWST NIRCam (blue squares with error bars; \citealt{weibel_2024}) and JWST COSMOS-Web (purple triangles with error bars; \citealt{shuntov_2025}).}
        \label{fg:mstar_density}
    \end{figure}

    Figure \ref{fg:merger_rates} shows the average stellar mass accretion rate density from major mergers, $\Psi_{\rm mer}$, selected with mass ratio $M_{\rm \star,sat} / M_{\rm \star,cen} > 0.25$ and central galaxy stellar mass $10^8 \, {\rm M_{\odot}} < M_\star < 10^{10} \, {\rm M_{\odot}}$. We compare simulations' predictions to the results from \citet{duan_2025} based on the JWST pair fractions and the TNG300 merger observability timescales. We also show for comparison the estimates obtained by integrating the specific mass accretion rate $\dot{M}_\star / M_\star$ from \citet{puskas_2025} and \citet{calabro_2026} along with the observed SMF $\phi(M_\star)$ from \citet{navarro_carrera_2024} and \citet{harvey_2025}
    \begin{equation}
        \Psi_{\rm mer} = \int_{\log M_{\rm \star, min} }^{\log M_{\rm \star, max}} \frac{\dot{M}_\star}{M_\star} \cdot \phi(M_\star ) \cdot M_\star \, \mathrm d \log M_\star \, ,
    \end{equation}
    in the stellar mass and redshift ranges of interest. We find that, overall, the merger accretion rate density evolutions predicted from cosmological models and simulations are broadly consistent both in normalisation and redshift-dependence with that measured by \citet{duan_2025}, \citet{puskas_2025} and \citet{calabro_2026}, with only \textsc{L-Galaxies} presenting a steeper increase in cosmic time which implies a slightly more rapid growth from mergers.

    \begin{figure*}
        \centering
        \includegraphics[width=0.8\textwidth]{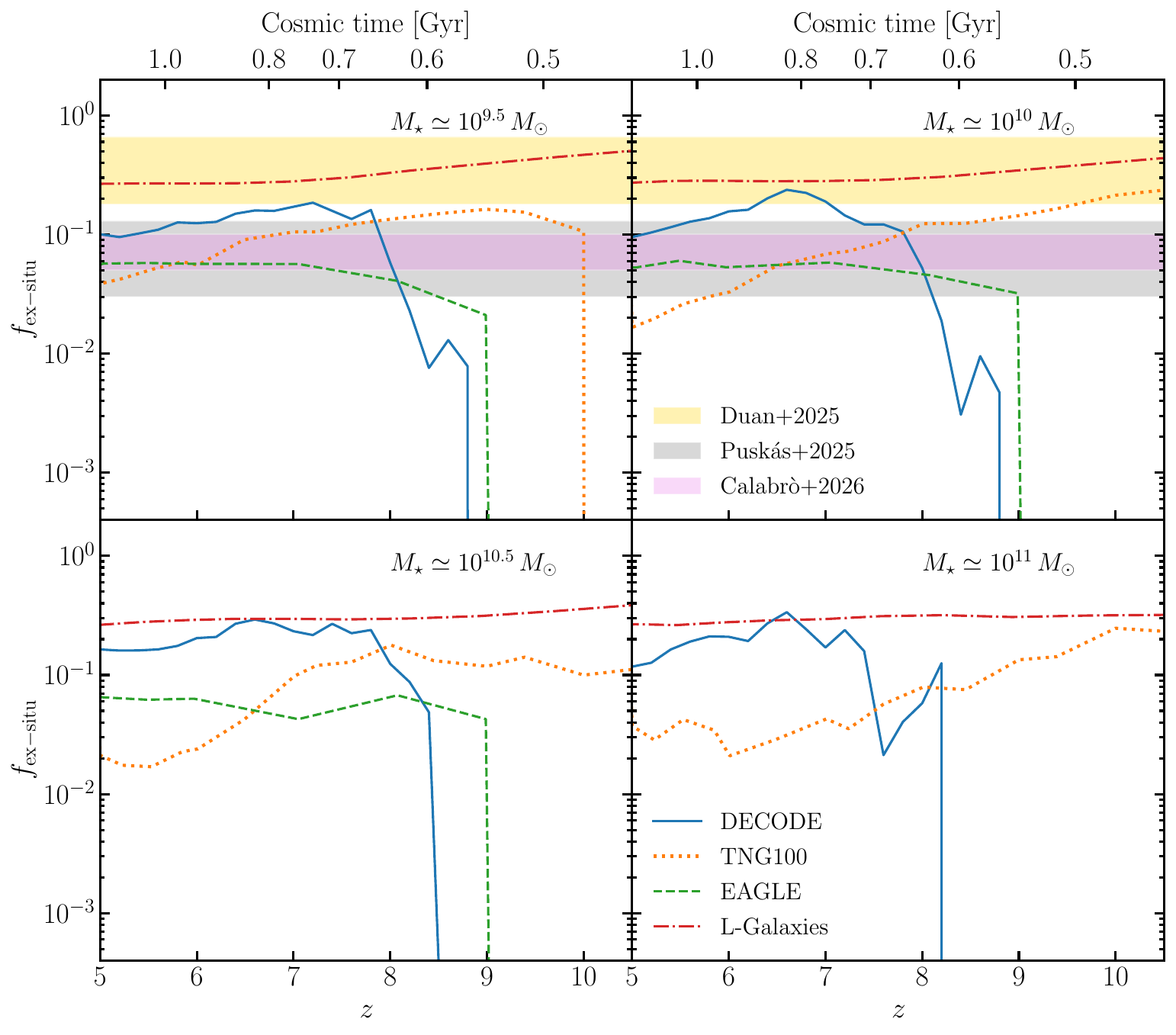}
        \caption{Average ex-situ fraction of accreted stellar mass as a function of redshift, for galaxies selected in different stellar mass bins at $z=5$. Galaxies are selected in stellar mass bins of $0.2$ dex width. The results from the TNG simulation (orange dotted lines), EAGLE simulation (green dashed lines) and \textsc{L-Galaxies} semi-analytical model (red dash-dotted lines) are also shown for comparison. The yellow, grey and pink shaded areas show the ex-situ fractions observationally inferred from \citet{duan_2025}, \citet{puskas_2025} and \citet{calabro_2026}.}
        \label{fg:frac_exsitu}
    \end{figure*}

    %\rezar{This is a quite useful addition. Could we also give a little more detail on how these points were obtained? \citet{harvey_2025} covers $6.5<z<13.5$, while Figure \ref{fg:merger_rates} has reconstructed points at $z\simeq5-6$. Was the SMF extrapolated there, or did we use another dataset? Or maybe I'm missing sth? Are we assuming a mass-independent specific accretion rate within the integration range?  I'd call these reconstructed estimates in the caption, since their normalizations also depend on our adopted SMF.}

    \subsection{Fraction of ex-situ stellar mass}\label{sec:res_exsitu}

    We now discuss the fraction of stellar mass that galaxies form via mergers against star formation. Figure \ref{fg:mstar_density} shows the SMD from mergers (ex-situ contribution only). We compare the evolution of the merger-driven SMD gathered from the \decode model, EAGLE and TNG simulations, and \textsc{L-Galaxies} semi-analytical model to the total SMD from a range of datasets in the literature such as COSMOS2020 (\citealt{weaver_2023}) and JWST (\citealt{navarro_carrera_2024, weibel_2024, shuntov_2025}). We find that all models suggest that the merger SMD evolution is one order of magnitude lower compared to the total cosmic SMD, and up to two orders of magnitude lower at $z\gtrsim 8$, supporting an in-situ star formation-dominated scenario.

    Figure \ref{fg:frac_exsitu} shows the ex-situ stellar mass fraction evolution for galaxies selected in different stellar mass bins at redshift $z=5$, predicted by \decode, TNG100, EAGLE and \textsc{L-Galaxies}. The ex-situ fraction is defined as the cumulative stellar mass growth from mergers normalised to the total stellar mass assembly. Observationally inferred ex-situ fractions from \citet{duan_2025}, \citet{puskas_2025} and \citet{calabro_2026}, for galaxies selected in their sample at each redshift, are also plotted for comparison. The average fraction of stellar mass accreted ex-situ via mergers is confined to $\lesssim 10-25\%$, depending on the galaxy stellar mass bin, with the only exception of \textsc{L-Galaxies} in the highest redshift bins. In particular, \decode's ex-situ fraction is relatively low at $z>8$, with a relatively sudden increase at $z\sim 8$, reaching $10-20\%$ and remaining relatively constant down to $z\sim 5$. We also verified that these results are insensitive to the adopted initial conditions for the stellar mass of central galaxies. Instead, other models suggest a relatively constant value of $\sim 10\%$ for TNG and EAGLE, and $\sim 25\%$ for \textsc{L-Galaxies}. At redshift $z\gtrsim5-6$, all models and simulations report an ex-situ fraction of $\lesssim25\%$, supporting the view that the stellar mass growth is mostly driven by in-situ star formation in the very early Universe.

\section{Discussion}\label{sec:discuss}

\subsection{Merger timescales and numerical uncertainties}

In this work, we performed a comprehensive study of the galaxy merger rates at high redshifts, employing cosmological semi-empirical models, hydrodynamical simulations and semi-analytical models. As mentioned above, measuring galaxy merger rates is challenging from the observational point of view. Several observations of mergers are carried out via galaxy pairs on typical projected separations of 5 to 50 kpc, which are assumed to merge within some assumed timescales (e.g., \citealt{mundy_2017, duan_2025}) typically derived from cosmological simulations. These inferred rates are therefore sensitive to the assumptions about orbital parameters, mass ratios, environmental effects and resolution of the simulations. For example, smaller galaxies are more prone to produce premature or false mergers (over-merging) in low-resolution simulations. This effect is especially relevant in large-volume simulations such as TNG300, where the mass resolution limits the reliable identification of low-mass satellites. To mitigate these issues, the semi-empirical model \decode anchors the galaxy merger timescales to the evolution of the dark matter subhaloes (Figures \ref{fg:surv_shmf} and \ref{fg:ushmf}), and analytically traces the full assembly histories of central galaxies with virtually no resolution issue (see, e.g., Figure \ref{fg:total_unevolved_shmf}).

Since \decode's constraints are based on the results of dark matter-only simulations, our inferred ex-situ fractions represent conservative estimates. Indeed, baryonic processes such as tidal stripping and dynamical friction, respectively, can shorten or prolong the merger event with respect to their host or even prevent coalescence entirely, depending on the mass of the host halo. This approach yields merger timescales that are systematically a factor $\sim 30\%$ longer/shorter, depending on the host halo mass, than those commonly computed in hydrodynamical simulations such as in TNG100 and TNG300, which are more consistent with their dark matter-only counterparts. This may be because, in the presence of baryons, the enhanced central densities of baryons may create higher resistance to tidal disruption effects in more massive haloes, whereas they accelerate orbital decay through dynamical friction and tidal forces in less massive haloes, therefore leading to systematically slightly different merger timescales. However, this variation in merger timescale alters our resulting galaxy merger rates and ex-situ fractions, we checked, by a few percent on average, and does not influence our results qualitatively.

\subsection{Comparisons with the literature}

Our predicted ex-situ fractions of $\lesssim10-25\%$ are broadly consistent with the predominantly in-situ growth inferred by \citet{puskas_2025} and \citet{calabro_2026}. Combining JADES close-pair measurements with a model of galaxy growth, \citet{puskas_2025} inferred cumulative major merger contributions increasing from approximately $3\%$ at $z=9$ to $13\%$ at $z=3$, using merger rates calibrated over $M_\star=10^8-10^{10}\,M_\odot$. Similarly, \citet{calabro_2026} found merger-driven stellar mass accretion rates approximately $5-10\%$ at $5<z<10$, including mergers with mass ratios down to $\mu=0.1$. High close-pair fractions can coexist with the predominantly in-situ growth found here, since the mergers contribution from observational works depends on merger probabilities, observability timescales, companion stellar masses, and the competing growth through star formation (e.g., \citealt{snyder_2017, puskas_2025, calabro_2026}). Galaxies that appear close on the sky may lie at different distances, but still be counted as candidate pairs if their photometric redshift estimates overlap within their uncertainties. Although probabilistic methods account for these uncertainties, some chance alignments may remain and systematically increase the inferred pair fraction (e.g., \citealt{puskas_2025}).

In contrast, our predicted fractions are substantially lower than the merger contribution inferred by \citet{duan_2025}. This apparent discrepancy requires accounting for differences in definition and sample completeness. The difference in adopted stellar mass ranges is not the main cause of this discrepancy, because restricting our analysis to $8<\log_{10}(M_\star/M_\odot)<10$ changes the inferred SMD by less than $0.045$ dex. Instead, the absence of corrections for galaxies below the observational limit in \citet{duan_2025}'s SFR density estimate provides a plausible contribution to its lower normalisation and potentially to the higher inferred relative importance of mergers. Quantifying this effect, however, requires a consistent treatment of completeness for both the merger accretion and star formation channels.

Independent support for predominantly in-situ assembly comes from the THESAN-ZOOM radiation-hydrodynamic simulations. \citet{shen_2026} found in-situ fractions close to unity for most haloes with $M_{\rm halo}\lesssim10^{11}\,M_\odot$ at $z\simeq6$ and $10$, although individual systems exhibit substantial scatter and mergers become more important in massive haloes at $z\simeq3$. Further support comes from the COLIBRE simulations, in which \citet{chaikin_2026} studied simulated counterparts of six luminous JWST galaxies at $z\gtrsim10$ and their evolutionary histories. The median in-situ stellar mass fractions exceed $75\%$ at all redshifts examined, although individual galaxies can exhibit slightly larger ex-situ contributions.

\subsection{Implications for galaxy growth}

A small ex-situ stellar mass fraction does not exclude an indirect contribution of mergers to galaxy growth, since interactions can drive gas inflows and enhance star formation within the wet mergers. \citet{calabro_2026} and \citet{duan_2026} reported enhanced SFRs in morphologically selected mergers and close pairs, respectively, although the latter find no statistically significant enhancement in specific SFR. \citet{puskas_2026} found a modest specific SFR enhancement of approximately $10\%$, whose strength depends on the timescale over which star formation is measured. These results suggest that interaction-induced star formation depends on the sample selection and temporal scales. Consistently, \citet{fu_2026} showed that intense star formation in the progenitors of massive high-redshift galaxies coincides with frequent major mergers at \(z\gtrsim9\), suggesting that mergers may promote early stellar mass growth with their gas content, despite the contribution in stellar mass from the stars they brought is low. In \decode, such effects are implicitly reflected in the empirically inferred SFRs and halo accretion histories and merger trees, although their individual contributions are not explicitly isolated. Stars formed within the main progenitor remain classified as in-situ regardless of the triggering mechanism and, consequently, predominant in-situ growth is compatible with a role for galaxy interactions.

A key implication of our results is that the lack of mass dependence of the major merger rates of dark matter haloes (e.g., \citealt{fakhouri_2008, fakhouri_2010, genel_2010}) translates also into a moderate dependence for the merger rates of galaxies (in line with the results from the Illustris simulation of \citealt{rodriguez_gomez_2015}). This is a direct consequence of the single-slope shape of the SMHM relation at high redshift, as we showed in \citetalias{fu_2026}. Instead, in the late Universe the SMHM relation is characterised by a pronounced double power law shape which generates increasing galaxy merger rates as a function of stellar mass, although the halo merger rates remain constant in halo mass.

\section{Conclusions}\label{sec:conclusions}

In this paper, we have investigated the merger histories of galaxies at $z \gtrsim 5$ by combining complementary theoretical approaches within the hierarchical $\Lambda$CDM framework, including the \decode semi-empirical model, the TNG and EAGLE cosmological hydrodynamical simulations, and the \textsc{L-Galaxies} semi-analytical model. By comparing predictions across these independent theoretical approaches, while also exploiting dark matter-only simulations to constrain merger timescales in \decode, we have constructed a comprehensive picture of the role of mergers in the early stellar mass assembly of galaxies. Despite the different physical assumptions and numerical techniques adopted in these models, they all tend to converge to the same conclusion that galaxy stellar mass growth at high redshifts is dominated by in-situ star formation, with mergers contributing only a relatively small fraction of the total stellar mass growth.

The main findings of this paper are summarised as follows:
\begin{itemize}
    %\item The Millennium and TNG100-Dark dark matter-only simulations predict merger timescales approximately $\sim 30\%$ longer than those from cosmological hydrodynamical simulations (TNG100 and TNG300) for haloes with mass $M_{\rm h} < 10^{12} \, M_\odot$ (Figure \ref{fg:tau_merge}).
    \item We provide the analytic fits of merger timescales that faithfully reproduce the statistics of unmerged surviving subhaloes at $z\simeq 4-9$ (Figures \ref{fg:surv_shmf} and \ref{fg:tau_merge}) and are comparable to those measured in hydrodynamical simulations within $\sim 30\%$, attributable to baryonic effects.
    \item Dark matter halo merger rates steadily increase with redshift at $z\gtrsim5$, nearly independent of the host halo mass (Figure \ref{fg:DMhalo_merger_rate}).
    \item Galaxy major merger ($M_{\rm \star,sat} / M_{\rm \star,cen} > 0.25$) rates in \decode are relatively constant in redshift  for the less massive galaxies, peaking at $\sim 3/{\rm Gyr}$ at $z\sim 7$ and decreasing with redshift for $M_\star > 10^{11}\, M_\odot$. Hydrodynamical simulations suggest a major merger rate ranging between $\sim 0.1/{\rm Gyr}$ and $\sim 5/{\rm Gyr}$ at $z=5-11$, with EAGLE showing a mild decrease at $z\gtrsim 7-8$, and TNG gradually increasing with redshift. \textsc{L-Galaxies} reports slightly higher major merger rates of a few tens per ${\rm Gyr}$ for $M_\star > 10^{10}\, M_\odot$ and $\sim 10/{\rm Gyr}$ in less massive galaxies (Figure \ref{fg:majormerger_rates_perGal}).
    \item All the cosmological models and simulations explored in this work predict that the total accretion rate density from mergers decreases with redshift (Figure \ref{fg:merger_rates}). They also suggest that mergers contribute $\lesssim 10\%$ at $z\lesssim 8$, and only a few percent at $z\gtrsim 8$, to the total cosmic stellar mass density (Figure \ref{fg:mstar_density}).
    \item The ex-situ fraction is confined to $\sim {\rm few} -25\%$ at $z\gtrsim5$, being broadly consistent among the cosmological models and simulations considered (Figure \ref{fg:frac_exsitu}), with differences possibly due to baryonic effects, implying an early growth of galaxies largely dominated by in-situ star formation.
\end{itemize}

In closing, our findings point to low stellar mass growth from mergers at high redshift. Our results have important implications for the interpretation of recent JWST observations, where high pair fractions have been taken as evidence for merger-driven growth. Our findings, instead, suggest that a significant fraction of these pairs may not merge efficiently within the short time available at high redshift or some of them may also be projected coincidences. Future observations combining pair statistics and resolved morphologies from JWST and upcoming wide-area surveys will provide important tests of this scenario by directly probing the dynamical state. In particular, measurements of close-pair survival and disturbed morphologies will help determine whether galaxy assembly during cosmic dawn is primarily driven by mergers or by sustained in-situ star formation fuelled by rapid gas accretion.

%\rezar{A few small points for the reply: ``like in Duan+25'' probably means ``unlike'', since they do provide a total mass accretion rate density, and ``retain this additional figure to keep the manuscript more concise'' seems to mean ``omit''. Otherwise, the reply is clear and helpful.}

%\rezac{Finally, the abstract puts the merger-rate peak at $z\simeq7$, while the Conclusions still say $z\simeq8$; please check which matches the updated figure.}

\begin{acknowledgments}
The authors thank the anonymous referee for their thorough review of the manuscript and comments. We thank Q. Duan for useful discussion on the observed galaxy pair abundances and ex-situ fractions. HF acknowledges support at Fudan University from the Shanghai Super Post-doctoral Excellence Program grant No. 2024008 and from the National Natural Science Foundation of China (NSFC) RFIS grant No. W2533002. MA is supported at the Argelander Institut f\"ur Astronomie through the Argelander Fellowship. AL has been supported by the Istituto Nazionale di Fisica Nucleare (INFN) via the specific national initiative QGSKY. FY is supported by the NSFC grants 12192220, 12192223 and 12133008. ALM acknowledges support through the European Space Agency (ESA) Research Fellowship in Space Science. EC acknowledges support from STFC consolidated grant ST/X001075/1. YP acknowledges support from the National Natural Science Foundation of China (NSFC) under grant Nos. 12125301 and 12192222, and from the New Cornerstone Science Foundation through the XPLORER PRIZE. KW is supported by the Science and Technology Facilities Council (STFC) through grant ST/X001075/1.
\end{acknowledgments}

\appendix

\section{Subhalo statistics}\label{app:shmf}

In this appendix we check that the unevolved total SHMF from SatGen fitted for satellites accreted at redshift $z_{\rm acc} < 1-2$, matches the total unevolved SHMF extracted from the Millennium-II and TNG100-Dark dark matter-only simulations.

\begin{figure*}
    \centering
    \includegraphics[width=0.9\textwidth]{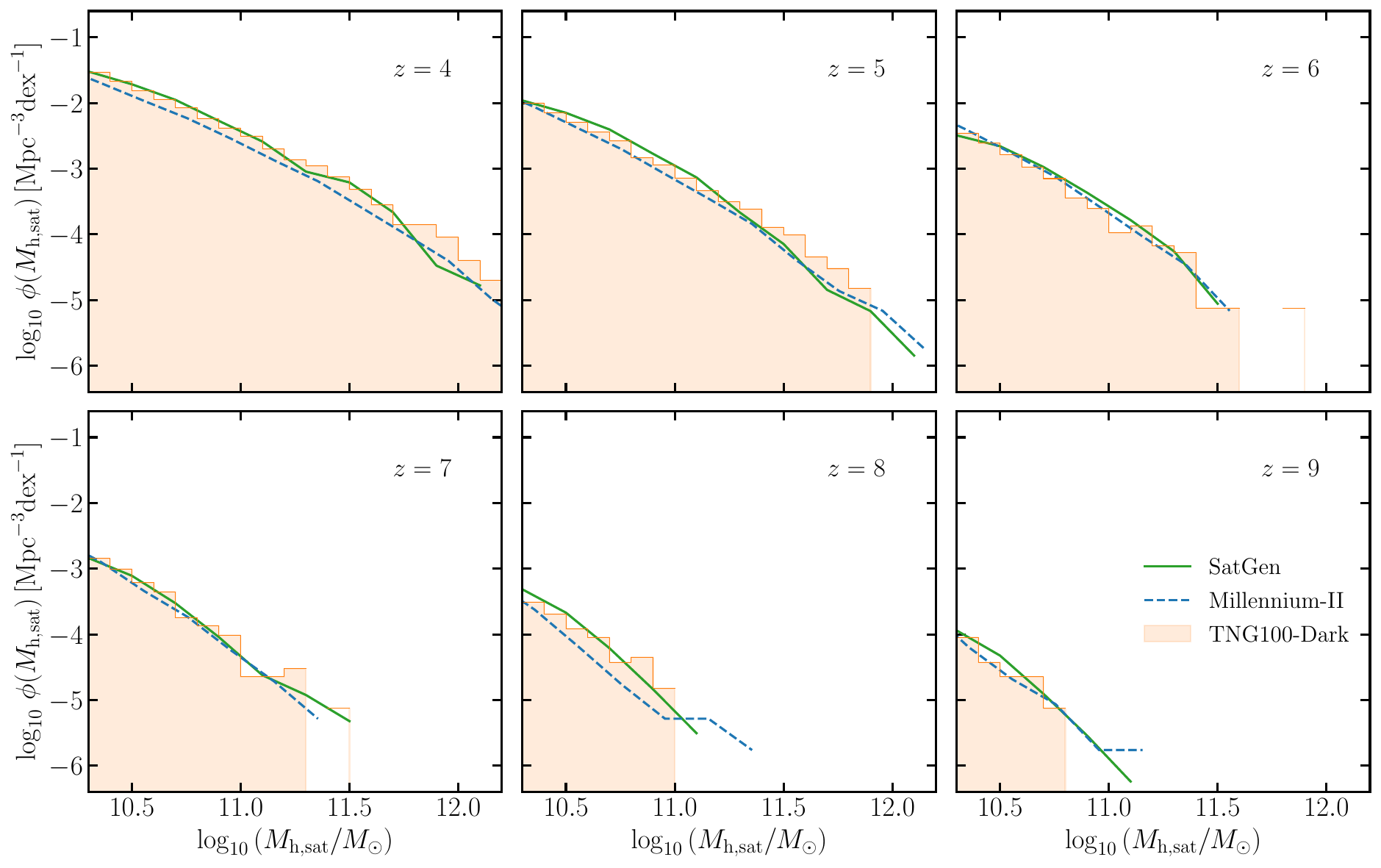}
    \caption{Unevolved total subhalo mass function at different redshifts, normalised to the cosmological volume. The green solid, blue dashed lines and orange shaded areas show the subhalo mass function from SatGen, Millennium-II and TNG100-Dark, respectively.}
    \label{fg:ushmf}
\end{figure*}

Figure \ref{fg:ushmf} shows the total unevolved SHMF from SatGen, used as input in the \decode semi-empirical model, compared to the one extracted from Millennium-II and the TNG100-Dark simulations. We rescaled the halo masses and volumes in the Millennium-II and SatGen, as they adopt slightly different cosmological parameters, to make the comparison consistent. The resulting unevolved SHMFs are in good agreement. We found that our subhalo statistics are in excellent agreement with those derived from dark matter-only simulations, at least above the mass resolution of the latter.

\begin{figure*}
    \centering
    \includegraphics[width=0.82\textwidth]{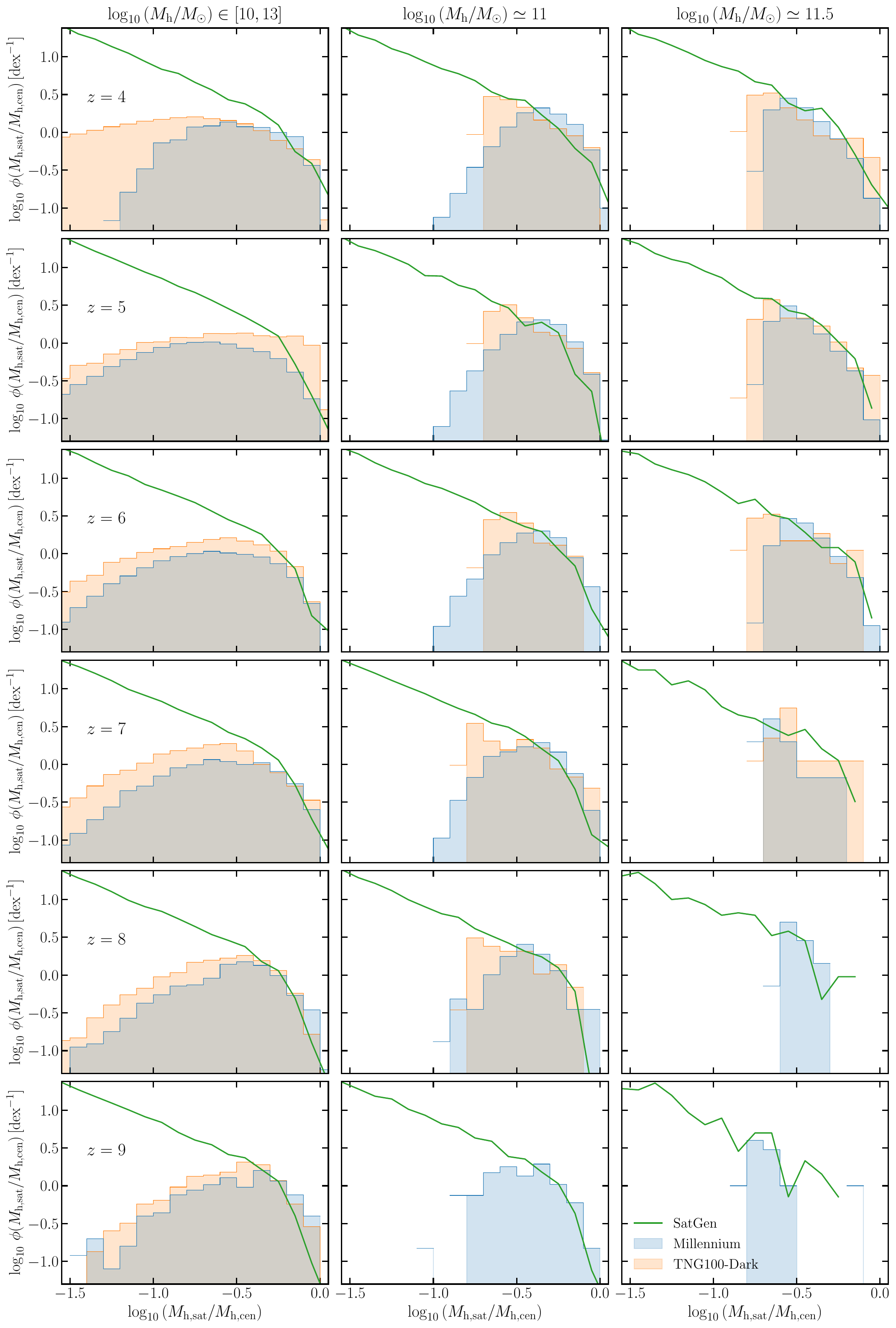}
    \caption{Total unevolved subhalo mass function at different redshifts normalised to the number of host haloes. Each column corresponds to one halo mass bin, as labelled. The green solid lines, blue and orange shaded areas show the subhalo mass function computed from SatGen, Millennium and TNG100-Dark, respectively.}
    \label{fg:total_unevolved_shmf}
\end{figure*}

We also show in Figure \ref{fg:total_unevolved_shmf} the same unevolved SHMF binned in halo mass ratio for different central halo mass bins ($10 < \log_{10} \, (M_{\rm h}/{\rm M_{\odot}}) < 13$, $\log_{10} \, (M_{\rm h}/{\rm M_{\odot}}) \simeq 11$ and $\log_{10} \, (M_{\rm h}/{\rm M_{\odot}}) \simeq 11.5$). The SHMF defined in this way provides insights on how the distribution of the subhaloes varies depending on the host halo mass bin. We report our mass functions compared to those computed from the Millennium and the TNG100-Dark simulations, showing relatively good agreement in all the halo mass bins from redshift $z=4$ to $z=9$. Our unevolved SHMF is also relatively well consistent with the universal fitting form from \citet{jiang_2014, jiang_2016}.

\bibliographystyle{aasjournalv7.1}
\bibliography{main}{}



\end{document}